\documentclass[aps, prb, twocolumn]{revtex4-1}

\usepackage {graphicx}
\usepackage {amsmath}
\usepackage {amsfonts}
\usepackage {amssymb}
\usepackage {mathrsfs}
\usepackage {bm}

\usepackage {epstopdf}
\allowdisplaybreaks[4]

\newcommand {\be}{\begin{eqnarray}}
\newcommand {\ee}{\end{eqnarray}}
\newcommand {\ba}{\be \begin{aligned}} 
\newcommand {\ea}{\end{aligned} \ee}
\newcommand {\bk}{\mathbf{k}}

\begin{document}

\title{Excitonic and superconducting orders from repulsive interaction on the doped honeycomb bilayer}



\author{James M. Murray}
\author{Oskar Vafek}
\affiliation{National High Magnetic Field Laboratory and Department of Physics, Florida State University, Tallahassee, FL 32306, USA}


\date{\today}

\begin{abstract}
\noindent 	Using a weak-coupling renormalization group formalism, we study competing ordered phases for repulsively interacting fermions on the bilayer honeycomb lattice away from half-filling, which can be realized experimentally as doped bilayer graphene. As electrons are added to the system, excitonic order is suppressed, and unconventional superconductivity appears generically in its place. In general it is found that the maximum critical temperature for superconductivity appears directly adjacent to the dome of particle-hole order, illustrating the importance of fluctuations in these channels for the formation of unconventional superconductivity. We obtain the phase diagram showing characteristic ordering temperatures for both short- and long-ranged interactions, and show that the most likely superconducting instabilities occur in $d$-wave, $f$-wave, and pair density wave channels. The nature of and competition between these phases are further analyzed using both free energy expansion and self-consistent mean-field theory. The effects of finite temperature and trigonal warping due to further-neighbor hopping are studied, and implications for experiments on bilayer graphene are discussed.
\end{abstract}


\maketitle


The formation of and competition between different types of ordered phases---and in particular between excitonic\cite{halperin68} (or ``particle-hole'') and superconducting phases---is a central issue in our quest for understanding quantum many-body physics. The idea that superconductivity can arise from repulsive interactions has a long history, dating back to the pioneering work of Kohn and Luttinger showing that the effective interaction between electrons in a metal can be attractive for channels with nonzero angular momentum, leading to the formation of Cooper pairs, even in cases where the bare electron interaction is entirely repulsive.\cite{kohn65} Despite providing an explicit mechanism for unconventional superconductivity in weakly coupled systems, the Kohn-Luttinger theory alone is unable to explain the empirically well-established fact that such unconventional superconductivity very often appears in close proximity to a phase with particle-hole order, and that systems that feature such competing phases tend to exhibit the highest superconducting transition temperatures. While it is widely believed that this proximity is not merely a coincidence, and that spin fluctuations or other soft modes from the nearby particle-hole phase tend to enhance superconductivity, there is so far no consensus regarding the precise mechanism by which this occurs.\cite{abanov01,honerkamp02,metlitski10,scalapino12,efetov13}

The bilayer honeycomb lattice in many ways provides an ideal arena in which to explore these questions. The rich band structure of this system, which features a high degree of symmetry, leads to the possibility of instabilities to many types of ordered phases. For the simplest case, in which only nearest-neighbor hopping of electrons is considered, the low-energy spectrum consists of two pairs of upward- and downward-dispersing parabolic bands touching at the charge neutrality point, with one pair each at the $\pm \mathbf{K}$ points of the Brillouin zone (for illustration, see Figure \ref{bilayer_dispersion}). In addition to being of purely theoretical interest, the bilayer honeycomb lattice has a physical incarnation as bilayer graphene, which can be readily studied experimentally. Experimental studies on suspended samples have shown evidence for the formation of interaction-driven symmetry breaking phases, with evidence emerging for both gapped \cite{velasco12,bao12} and gapless \cite{weitz10,feldman09,mayorov11} behavior at low energies. The fact that electron interactions in bilayer graphene are strong enough to lead to nontrivial many-body behavior while still being small enough to allow for the use of weakly-coupled theoretical approaches---as evidenced by the small energy scales ($\sim$ meV) at which ordering behavior has been seen experimentally---provides hope that the electronic properties of this material can be studied and understood both experimentally and theoretically. Theoretical studies of bilayer graphene using a variety of methods have led to many different possibilities for the ground state at the charge neutrality point, with proposals including layer polarized \cite{zhang10,nandkishore10}, nematic \cite{lemonik10,vafek10b}, antiferromagnetic \cite{castro_neto09,kharitonov12}, and quantum anomalous Hall \cite{nandkishore10b}.


The renormalization group (RG) is an attractive option for addressing behavior in systems with many competing phases, due in particular to the fact that---unlike standard mean-field theory---it is an unbiased approach that treats all types of order on an equal footing. Due to the fact that electron interactions in this system are marginally relevant, RG can be used to investigate ordering phenomena for arbitrarily weak values of the interaction strength. Recently we have shown in the context of an idealized model that unconventional superconductivity can be realized on the honeycomb bilayer by including a nonzero chemical potential \cite{vafek13}. Here we generalize that work by investigating the effects of nonzero temperature and trigonal warping due to further neighbor hopping, and we also resolve the nature of the superconducting phase, which requires analysis beyond the RG in cases where the leading instability corresponds to a two-dimensional space group representation (as occurs, for example, for $d$-wave superconductivity, which features degenerate order parameter components with $d_{x^2-y^2}$ and $d_{xy}$ symmetry, or, as explained in Section \ref{sec:SC}, for pair density wave superconductivity).

\begin{figure}
\centering
\includegraphics[width=0.45\textwidth]{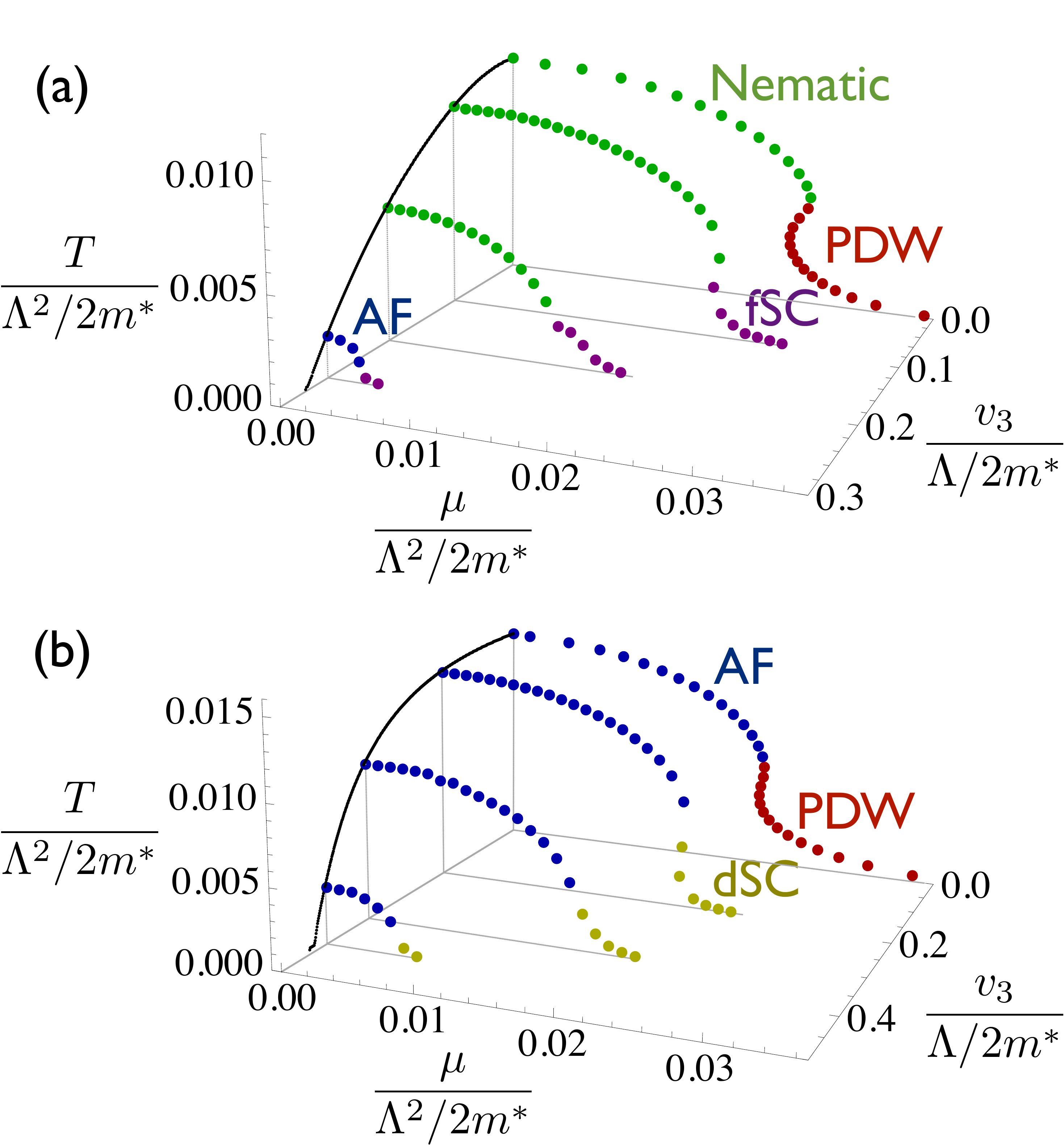}
\caption{Characteristic ordering temperatures as a function of chemical potential and the velocity $v_3$ associated with trigonal warping. (a) Long-ranged interactions near the forward-scattering limit, with bare couplings $g_{A_{1g}} = 0.15$ and $g_{A_{2u}} = g_{E_\mathbf{K}} = 0.003$, lead to nematic or antiferromagnetic instabilities near half-filling, giving way to pair density wave (PDW) or $f$-wave superconducting instabilities as $\mu$ is increased. (b) Short-ranged Hubbard interaction, with bare couplings $g_{A_{1g}} = g_{A_{2u}} = 0.06$ and $ g_{E_\mathbf{K}} = 0.03$, leads to antiferromagnetic, PDW, or $d$-wave superconducting instabilities. The cutoff energy and trigonal warping velocity for bilayer graphene are given by $\Lambda^2/2m^* \approx 0.2$ eV and $v_3 \approx 0.178 \Lambda/2m^*$, respectively.
\label{fig:phases3d}}
\end{figure}

\begin{figure}
\centering
\includegraphics[width=0.45\textwidth]{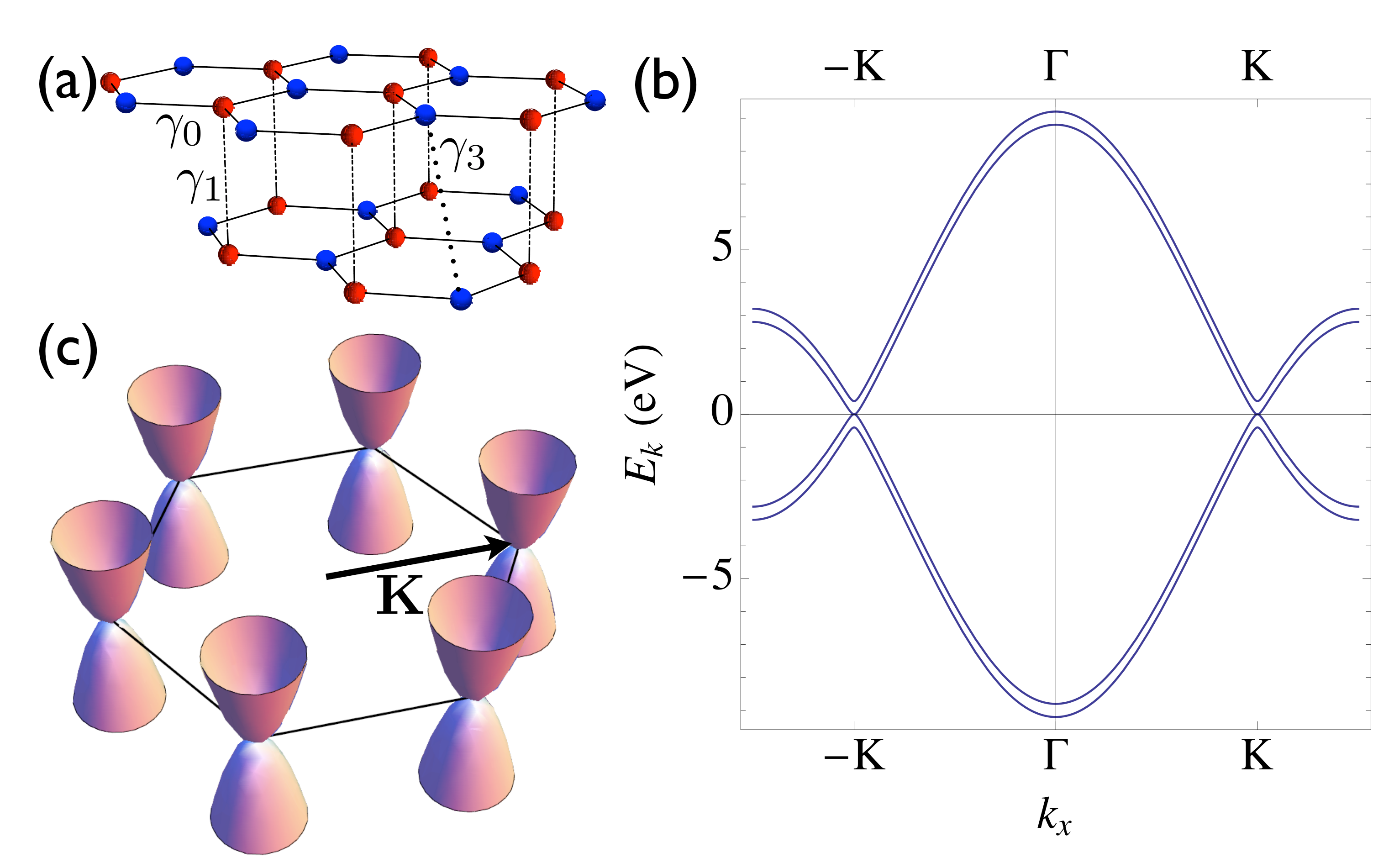}
\caption{(a) The AB stacked bilayer honeycomb lattice, with $\gamma_i$ corresponding to hopping between various sites. (b) Dispersion of bilayer graphene in the absence of trigonal warping, with parameters\cite{zhang08} \hbox{$\gamma_0=3.0$ eV}, \hbox{$\gamma_1=0.4$ eV}, and $\gamma_3=0$. (c) Dispersion in the low-energy effective theory, with parabolic bands touching at the Brillouin zone corner at $\mathbf{K} =  (\frac{4\pi}{3\sqrt{3}a},0)$.
\label{bilayer_dispersion}}
\end{figure}

Our main results obtained from solving the RG equations and analyzing susceptibilities are summarized in Figure \ref{fig:phases3d}, which shows the characteristic ordering temperatures for various phases as a function of the chemical potential $\mu$ and the velocity $v_3$ associated with trigonal warping, which distorts the low-energy parabolic spectrum.
For interactions of long (but finite) range, as illustrated in Figure \ref{fig:phases3d}(a), one obtains a nematic phase at small $\mu$, which gives way to a superconducting phase upon doping. For the case in which trigonal warping is absent or very small, the superconducting state is a pair density wave (PDW), in which the electrons form pairs within a pocket and so carry finite total momentum. For larger values of $v_3$, one obtains a spin-triplet $f$-wave superconductor, and eventually the nematic phase is replaced by antiferromagnetism. For short-ranged Hubbard interaction, on the other hand, one obtains the phases shown in Figure \ref{fig:phases3d}(b). The antiferromagnetic phase at small $\mu$ is suppressed by doping, leading again to a PDW phase for small $v_3$. For larger values of $v_3$, however, one obtains instead a $d$-wave superconducting phase, which, as we shall show, is chiral and breaks time-reversal symmetry. We emphasize that the bare electron-electron interactions used in obtaining Figure \ref{fig:phases3d} are entirely repulsive, with effective attractive interactions being generated through the RG flow. One can see from the figure that in all cases the maximum critical temperature for superconductivity is obtained directly adjacent to the excitonic phase, illustrating the importance of fluctuations in particle-hole channels for obtaining unconventional superconductivity. In the remainder of the paper we shall lay out the method used to obtain the results shown in Figure \ref{fig:phases3d}, providing a detailed analysis of the RG equations and associated phases.  

The outline of this paper is as follows. In Section \ref{sec:RG} we introduce the model and describe the RG procedure, which consists of deriving and solving coupled flow equations for chemical potential, temperature, trigonal warping velocity, and the nine symmetry-allowed fermion couplings. In the three sections that follow, these equations are solved for cases of increasing complexity. In Section \ref{sec:mu} we solve the RG equations at zero temperature and in the absence of trigonal warping, establishing the mechanism by which the combination of RG-generated attractive interaction and appropriately chosen chemical potential can lead to superconductivity. In Section \ref{sec:temperature} we introduce finite temperature, which tends to suppress the runaway flow of the couplings. The flows of the temperature and chemical potential are studied in the infrared limit, and we show that the system tends to condense in an excitonic or a superconducting phase, depending on which of these quantities is most relevant. In Section \ref{sec:trig} we introduce trigonal warping, which results from further-neighbor electron hopping and favors $d$- and $f$-wave superconductivity over the pair density wave state that is present without trigonal warping. In Section \ref{sec:SC} we use both free-energy expansion and self-consistent mean-field theory to complement our RG analysis and analyze the nature of the superconducting phase. It is found that the $d$-wave superconducting phase is chiral and breaks time-reversal symmetry, while the non-chiral phase is favored in the pair density wave case. Finally, in Section \ref{sec:discussion} we discuss the results, comparing with existing theories and commenting on the possible experimental implications of our results. Technical details concerning the flow equations and susceptibility analysis are provided in the appendices.

\section{Renormalization group procedure}
\label{sec:RG}
The low-energy Hamiltonian describing electrons on the honeycomb bilayer is $H = H_0 + H_\mathrm{int}$, with the noninteracting part given by \cite{nilsson08,vafek10}
\ba
\label{H0}
H_0 =& \sum_{|\mathbf{k}|<\Lambda} \sum_{\sigma=\uparrow,\downarrow}
	\psi^\dagger_{\mathbf{k}\sigma} \mathcal{H}_\mathbf{k} \psi_{\mathbf{k}\sigma} \\
\mathcal{H}_\mathbf{k} =& \frac{k_x^2 - k_y^2}{2m^*} 1\sigma_1 
	+ \frac{k_x k_y}{m^*} \tau_3\sigma_2 \\
	&\quad\quad+ v_3 k_x \tau_3\sigma_1 - v_3 k_y 1\sigma_2.
\ea
The Pauli matrices $\tau_i$ and $\sigma_i$ operate in valley and layer spaces, respectively, and $\sigma$ denotes the electron spin. Experimentally, the effective mass is given by\cite{mayorov11,velasco12} $m^* \approx 0.029m_e$, while the trigonal warping velocity, which distorts the parabolic low-energy spectrum into four Dirac cones near each of the points $\pm \mathbf{K} = \pm(\frac{4\pi}{3\sqrt{3}a},0)$, is given by\cite{mayorov11} $v_3 \approx 1.41\times 10^5$ m/s. The cutoff energy scale $\Lambda^2 / 2m^* \sim 0.2$ eV is roughly given by the splitting of the upper- and lowermost bands at $\pm \mathbf{K}$. 

The interacting part of the Hamiltonian is given by \cite{cvetkovic12}
\be
\label{H_int}
H_\mathrm{int} = \frac{2\pi}{m^*} \sum_{i=1}^9 g_i \sum_{m=1}^{m_i} \int d^2 x
	\left( \sum_\sigma \psi^\dagger_\sigma (x) \Gamma_i^{(m)} \psi_\sigma (x) \right)^2,
\ee
where $i$ is summed over the 9 irreducible representations of the lattice space group, which is D$_{3d}$ at $\mathbf{\Gamma}=(0,0)$, and D$_3$ at $\mathbf{K}$. There is a unique coupling $g_i$ corresponding to each representation, and $m_i$ denotes the multiplicity within a representation. The 16 interaction matrices $\Gamma_i^{(m)}$ are provided in Table \ref{table:matrices}. 
\begin{table*}
\centering
    \begin{tabular}{lllllll}
    \hline \hline
        Rep. & PH matrix $\Gamma_i^{(m)}$ & PH phase (charge) & PH phase (spin) & PP matrix $\tilde\Gamma_i^{(m)}$ & SC phase & SC spin\\ \hline
        $A_{1g}$ 	& $1_4$ & Charge instability & Ferromagnet & $\tau_1 1$ & $s$ & singlet \\
        $A_{2g}$ 	& $\tau_3 \sigma_3 $ & Anomalous quant.~Hall\cite{nandkishore10b,haldane88} & Quant.~spin Hall\cite{throckmorton12,lemonik10,scherer12} & $-\tau_2 \sigma_3$ & $f_\pm$ & triplet \\
        $E_g$ & $1 \sigma_1 , \tau_3 \sigma_2 $ & Nematic\cite{vafek10b,lemonik10} & Spin nematic & $\tau_1 \sigma_1, \tau_2 \sigma_2$ & $d_{x^2-y^2}, d_{xy}$ & singlet \\
        $A_{1u}$ & $\tau_3 1$ & Loop current\cite{zhu13} & Staggered spin current & $-\tau_2 1$ & $f$ & triplet \\
        $A_{2u}$ & $1 \sigma_3 $ & Layer polarized\cite{zhang10,min08,nandkishore10} & Layer AF\cite{castro_neto09,vafek10,kharitonov12} & $-\tau_1 \sigma_3$ & $s_\pm$ & singlet \\
        $E_{u}$ & $\tau_3 \sigma_1 , -1 \sigma_2 $ & Loop current II & Loop spin current II & $\tau_2 \sigma_1, \tau_1 \sigma_2$ & $p_x, p_y$ & triplet \\
        $A_{1\mathbf{K}}$ & $ \tau_1 \sigma_1 , \tau_2 \sigma_1 $ & Kekul\'{e}\cite{hou07} & Spin Kekul\'{e} & $1\sigma_1, \tau_3 \sigma_1$ & $s$-PDW & singlet\\
        $A_{2\mathbf{K}}$ & $\tau_1 \sigma_2 , \tau_2 \sigma_2 $ & Kekul\'{e} current & Spin Kekul\'{e} current & $-1\sigma_2 , \tau_3 \sigma_2$ & $p$-PDW & triplet \\
        $E_{{\mathbf K}}$ & $\tau_1 1, - \tau_2 \sigma_3 , - \tau_2 1, - \tau_1 \sigma_3 \quad$ & Charge density wave & Spin density wave & $1_4, \tau_3 \sigma_3 , \tau_3 1, 1\sigma_3 \quad$ & $d$-PDW & singlet \\
    \hline \hline
    \end{tabular}
    \caption{The 9 space-group representations of the bilayer honeycomb lattice and matrices appearing in the particle-hole (PH) and particle-particle (PP) fermion bilinears transforming according to each representation. The third and fourth columns give the names of the phases realized by condensation of the particle-hole bilinears in charge and spin channels.\cite{cvetkovic12} The sixth column gives the name of the superconducting state phase realized upon condensation of the particle-particle bilinear, where $\pm$ denotes a change of sign between layers and PDW denotes pair density wave states,and the final column lists the spin of the superconducting phase.}
    \label{table:matrices}
\end{table*}
The factor of $\frac{2\pi}{m^*}$ in \eqref{H_int} makes the couplings $g_i$ dimensionless. We shall work in units with $\hbar = k_B=1$.
Further insight can be gained by noting that the interaction term \eqref{H_int} can be rewritten as a sum of {\it particle-particle} interactions:
\ba
\label{H_pp}
H_\mathrm{int} =& \frac{2\pi}{m^*} \sum_{\substack{i=1,3,\\ 5,7,9}} \tilde g_i 
	\sum_{m=1}^{m_i} S_i^{(m)\dagger} S_i^{(m)} \\
	&+ \frac{2\pi}{m^*} \sum_{\substack{i=2,4,\\ 6,8}} \tilde g_i 
	\sum_{m=1}^{m_i} \mathbf{T}_i^{(m)\dagger} \cdot \mathbf{T}_i^{(m)} ,
\ea
where the couplings $\tilde g_i$ are related to the original couplings by a Fierz transformation\cite{vafek13}:
\begin{widetext}
\be
\left( \begin{matrix} 
	&\tilde  g_{A_{1g}} \\ 
	&\tilde  g_{A_{2g}} \\ 
	&\tilde  g_{E_{g}} \\ 
	&\tilde  g_{A_{1u}} \\ 
	&\tilde  g_{A_{2u}} \\ 
	&\tilde  g_{E_u} \\ 
	&\tilde  g_{A_{1\mathbf{K}}} \\ 
	&\tilde  g_{A_{2\mathbf{K}}} \\ 
	&\tilde  g_{E_\mathbf{K}}
	\end{matrix} \right)
= \frac{1}{8} \left(
   \begin{matrix} 
      1 & -1 & 2 & -1 & 1 & -2 & 2 & -2 & 4 \\
      1 & -1 & -2 & -1 & 1 & 2 & 2 & -2 & -4\\
      1 & 1 & 0 & -1 & -1 & 0 & 2 & 2 & 0 \\
      1 & -1 & 2 & -1 & 1 & -2 & -2 & 2 & -4 \\
      1 & -1 & -2 & -1 & 1 & 2 & -2 & 2 & 4 \\
      1 & 1 & 0 & -1 & -1 & 0 & -2 & -2 & 0 \\
      1 & -1 & 2 & 1 & -1 & 2 & 0 & 0 & 0 \\
      1 & -1 & -2 & 1 & -1 & -2 & 0 & 0 & 0 \\
      1 & 1 & 0 & 1 & 1 & 0 & 0 & 0 & 0 
   \end{matrix} \right)
\left( \begin{matrix} 
	&  g_{A_{1g}} \\ 
	&  g_{A_{2g}} \\ 
	&  g_{E_{g}} \\ 
	&  g_{A_{1u}} \\ 
	&  g_{A_{2u}} \\ 
	&  g_{E_u} \\ 
	&  g_{A_{1\mathbf{K}}} \\ 
	&  g_{A_{2\mathbf{K}}} \\ 
	&  g_{E_\mathbf{K}}
	\end{matrix} \right).
\ee
\end{widetext}
This new basis turns out to be more convenient for describing fluctuations and instabilities in superconducting channels, and the couplings $\tilde g_i$ shall be used in much of what follows. The singlet and triplet particle-particle bilinears in \eqref{H_pp} are defined as
\ba
S_i^{(m)} &= \sum_\bk \sum_{\alpha,\beta}
	\psi^\dagger_{\bk,\alpha} \tilde\Gamma_i^{(m)}(\sigma_2)_{\alpha\beta} 
	\psi^*_{-\bk,\beta} \\
\mathbf{T}_i^{(m)} &= \sum_\bk \sum_{\alpha,\beta}
	\psi^\dagger_{\bk,\alpha} \tilde\Gamma_i^{(m)}
	(i\sigma_2\boldsymbol{\sigma})_{\alpha\beta} \psi^*_{-\bk,\beta},
\ea
and the matrices $\tilde\Gamma_i^{(m)}$ are also given in Table \ref{table:matrices}.

In implementing the RG at finite temperature $T$, it is useful to define the following action:
\ba
\label{action}
S = \int_0^{1/T} d\tau \bigg\{ \sum_{|\mathbf{k}|<\Lambda,\sigma} 
	& \psi_{\bk\sigma}^\dagger [\partial_\tau + \mathcal{H}_\bk -(\mu+\delta\mu)] \psi_{\bk\sigma} \\
	&+ H_\mathrm{int} \bigg\},
\ea
where $\psi_{\bk\sigma} = \psi_{\bk\sigma} (\tau)$ are treated as Grassmann fields. The chemical potential of the half-filled system is nonzero in the presence of interactions and is denoted as $\delta\mu$. With this notation, $\mu$ is the deviation of the chemical potential away from half filling, with $\mu=0$ corresponding to the half-filled system. Details regarding the chemical potential, including the exact value of $\delta\mu$, and its RG flow are provided in Appendix \ref{sec:flow_coefficients}. 

\begin{figure}
\centering
\includegraphics[width=0.45\textwidth]{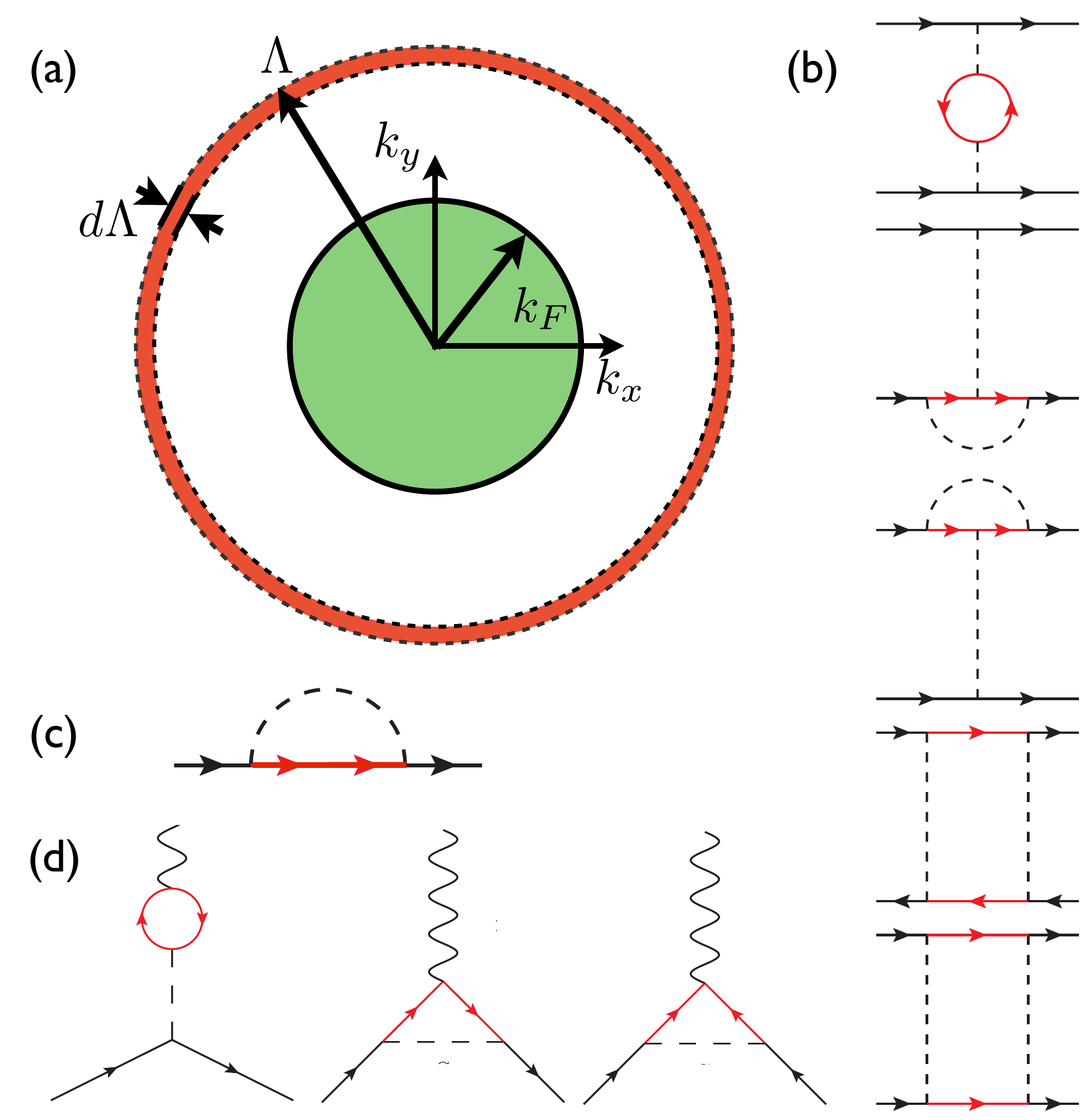}
\caption{Schematic depiction of the RG procedure. (a) Near momenta $\pm \mathbf{K}$, modes with large momentum (shown in red) near the UV cutoff $\Lambda$ are integrated out, leading to an effective theory with a smaller cutoff and modified parameters.  (b) Diagrams showing RG flow contributions to the fermion couplings $g_i$, where the red internal lines are integrated over ``fast modes'' with momentum near $\Lambda$. (c) Diagram showing the RG flow contribution to the chemical potential. (d) Diagrams showing RG flow contributions to the vertex functions in Equation \eqref{source_terms}. The first two diagrams correspond to particle-hole vertex, while the last corresponds to the particle-particle vertex.
\label{RG}}
\end{figure}

We begin the RG procedure by integrating out fermionic states within a shell of momenta $e^{-\ell} \Lambda < k < \Lambda$, as shown in Figure \ref{RG}, where $\ell>0$ is the RG flow parameter, while summing over all Matsubara frequencies for those states. After integrating out a momentum shell, we rescale the frequencies, momenta, fields, chemical potential, and trigonal warping velocity in such a way that the noninteracting part of the action \eqref{action} remains invariant. According to this tree-level rescaling, one finds that $T_\ell = T e^{2\ell}, \mu_\ell = \mu e^{2\ell}$, and $v_{3\ell} = v_3 e^{\ell}$. 
While the RG flows of $T$ and $v_3$ are not affected by one-loop corrections arising from interactions, the chemical potential flow is modified by such corrections. Taking into account the one-loop contribution due to interactions, one obtains the flow equation
\be
\label{mu_flow}
\frac{d \mu_\ell}{d\ell} = 2 \mu_\ell - 2 K (T_\ell, \mu_\ell, v_{3\ell}) \sum_i c_i g_i (\ell).
\ee
Note that some care is required in properly determining the flow of the chemical potential at nonzero temperature---see details in Appendix \ref{sec:flow_coefficients}. We also emphasize that the chemical potential is {\em not} held fixed under RG transformation, as in some other approaches\cite{polchinski92,shankar94}, but rather is treated as a relevant perturbation that grows upon running the RG. This is loosely analogous to the treatment of the mass parameter in the bosonic $n$-vector model\cite{chaikin95}, with the important difference that Cooper instabilities are not suppressed by finite $\mu$. The role of the chemical potential in the RG flow is further discussed in Section \ref{sec:discussion}. 

The couplings $g_i$, meanwhile, are marginal at tree level. They flow according to the following equation once one-loop interaction effects are included:
\be
\label{g_flow}
\frac{dg_i(\ell)}{d\ell} = \sum_{j,k=1}^9 \mathcal{A}_{ijk}(T_\ell, \mu_\ell, v_{3\ell}) g_j(\ell) g_k(\ell).
\ee
The diagrams leading to these equations are shown in Figure \ref{RG}.
The perturbative RG flow equations \eqref{mu_flow} and \eqref{g_flow} include all one-loop contributions and so provide a complete description of the interacting system in the weak-coupling limit $g_i \ll 1$.

We investigate possible types of symmetry breaking by introducing source terms into the Hamiltonian:
\ba
\label{source_terms}
H_\Delta &= \sum_{i=1}^{9} \Delta_i^{ph} \sum_{\bk,\alpha} 
\psi^\dagger_{\bk,\alpha} \Gamma_i^{(1)} \psi_{\bk,\alpha} \\
	&+ \sum_{i=10}^{18} \boldsymbol{\Delta}_{i}^{ph} \cdot \sum_{\bk,\alpha,\beta}
	 \psi^\dagger_{\bk,\alpha} \Gamma_i^{(1)} \boldsymbol{\sigma}_{\alpha\beta} 
	 \psi_{\bk,\beta} \\
	&+ \frac{1}{2} \sum_{\substack{i=1,3,\\5,7,9}} \left( \Delta_i^{pp} S_i^{(1)} + H.c. \right) \\
	&+ \frac{1}{2}\sum_{\substack{i=2,4,\\6,8}} 
	\left( \boldsymbol{\Delta}_i^{pp} \cdot \mathbf{T}_i^{(1)} + H.c. \right),
\ea
or, equivalently, adding $S_\Delta = \int d\tau H_\Delta$ to the action. The terms in the first and second summations in \eqref{source_terms} describe source fields coupling to the fermions in charge ({\em e.g.}\ nematic) and spin ({\em e.g.}\ antiferromagnetic) channels, respectively, while the third and fourth describe pairing in singlet and triplet channels, respectively. The classification of the pairing terms is shown in Table \ref{table:matrices}. The properties of the 18 particle-hole bilinears have been catalogued previously\cite{cvetkovic12} and are reproduced in Table \ref{table:matrices} for completeness. Because all components of a multi-dimensional representation have identical symmetry properties, only the first component of each representation ({\em i.e.}\ $\Gamma_i^{(m=1)}$ and $\tilde\Gamma_i^{(m=1)}$) is included in \eqref{source_terms}. Similarly, although the source terms in spin channels, $\boldsymbol{\Delta}_i^{ph,pp}$, each contain three components, $SU(2)$ symmetry dictates that there is no loss of generality in considering just one of these components ({\em e.g.}~the component coupling to the spin matrix $\sigma_3$). The particle-hole and particle-particle vertices $\Delta_i^{ph,pp}$  flow under RG according to the diagrams shown in Figure \ref{RG}(d), with
\be
\label{vertex_flow}
\frac{d \ln \Delta_i^{ph,pp}}{d\ell} 
	= 2 + \sum_{j=1}^9 B^{ph,pp}_{ij} (T_\ell, \mu_\ell, v_{3\ell}) g_j (\ell).
\ee
The explicit expressions for the coefficients $\mathcal{A}_{ijk}$, $K$, and $B^{ph,pp}_{ij}$ from \eqref{mu_flow}--\eqref{g_flow} and \eqref{vertex_flow} are provided in Appendices \ref{sec:flow_coefficients} and \ref{sec:bilayer_susceptibilities}. The flow equations \eqref{mu_flow}, \eqref{g_flow}, and \eqref{vertex_flow} can be easily solved numerically given an initial choice of couplings, temperature, chemical potential, and trigonal warping velocity. In the following three sections we shall study and solve the behavior of these equations for increasingly complex cases.

\section{Solution of flow equations with chemical potential}
\label{sec:mu}
In this section we solve the RG flow equations for the relatively simple case in which temperature and the trigonal warping velocity are set to zero. This shall allow us to clearly establish the mechanism by which unconventional superconductivity is realized in this system. The material in this section is partly a review of our previous results\cite{vafek13}, but is useful to recapitulate here as it shall provide context for the more complicated cases considered in the following sections, and will also form the basis for the mean-field study of the superconducting phases presented in Section \ref{sec:SC}.

For density-density interactions between electrons, only three of the nine couplings are nonzero initially: $g_{A_{1g}}$, $g_{A_{2u}}$, and $g_{E_\mathbf{K}}$, with all being positive for repulsive interaction. In general, one expects $g_{A_{1g}}$, which corresponds to forward scattering with small momentum transfer and hence long spatial range, to be the largest of these. As the spatial range of the interaction is decreased, the other two couplings play a greater role. In the Hubbard limit of on-site interaction, the couplings are all of the same order, with $g_{A_{1g}} = g_{A_{2u}} = 2g_{E_\mathbf{K}}$. If all three of these couplings are initially nonzero, then the six remaining couplings will all be generated under RG flow. In what follows we shall focus on the particular cases of short-ranged Hubbard and near-forward scattering interactions, with the bare couplings in the latter case given by $g_{A_{2u}} = g_{E_\mathbf{K}} = 0.02 g_{A_{1g}}$. (We do not present results for the pure forward scattering limit, in which only $g_{A_{1g}}$ is initially nonzero, as this fine-tuned case leads to flows along unstable trajectories and degenerate superconducting phase instabilities---see Figure 2 in Ref.~\onlinecite{vafek13}.) The resulting flows and phase instabilities for other choices of initial couplings are qualitatively similar to those presented for these two cases, so long as one considers repulsive density-density interactions.

\begin{figure}
\centering
\includegraphics[width=0.48\textwidth]{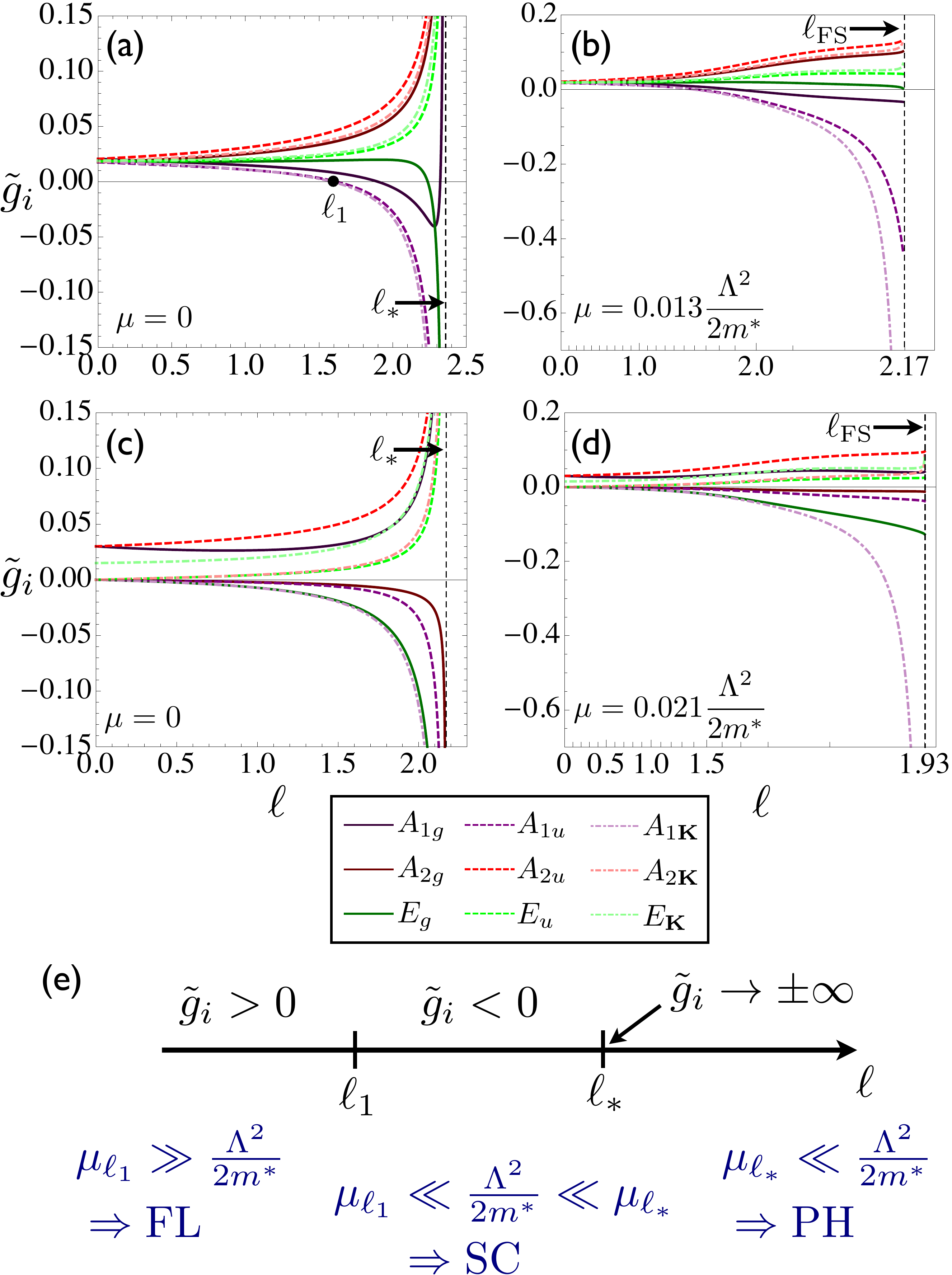}
\caption{RG flows of the couplings $\tilde g_i (\ell)$ for $T = v_3 = 0$, for various values of chemical potential. Upper panels are calculated for near-forward scattering interaction, with bare couplings $g_{A_{1g}} = 0.15$, $g_{A_{2u}} = g_{E_\mathbf{K}} = 0.003$. Middle panels are calculated for Hubbard interaction, with bare couplings $g_{A_{1g}} = g_{A_{2u}} = 0.06$, $ g_{E_\mathbf{K}} = 0.03$. Note that $\ell$ axis has been exponentially stretched near $\ell_\mathrm{FS}$ in (b) and (d) in order to show the behavior of the flows more clearly in this region. (e) Schematic behavior of the flows at $\mu=0$, showing the values of $\ell$ at which attractive interaction is generated and at which the couplings diverge, indicating a phase instability. The lower part shows the various behaviors that can be expected for different choices of chemical potential: Fermi liquid (FL), superconductivity (SC), and particle-hole order (PH).
\label{fig:g_flows0}}
\end{figure}

The solution to the flow equations for these two choices of repulsive bare interactions are shown in Figure \ref{fig:g_flows0}.
The solutions exhibit a number of generic features. The bare couplings $\tilde g_i (0)$ are all small and repulsive to begin with, except in the Hubbard limit, where some bare couplings vanish. Upon running the RG, some of these couplings become negative at $\ell = \ell_1$, indicating the potential for attractive pairing in some channel. (The Hubbard model is a special case, in which $\ell_1 = 0$). This attraction does not necessarily guarantee superconductivity, however, as repulsive couplings may also grow in magnitude under RG flow. For small values of chemical potential, as shown in Figure \ref{fig:g_flows0}(a,c), this is indeed the case, with attractive and repulsive couplings diverging as $\ell \to \ell_*$. As can be shown by computing susceptibilities, such behavior corresponds to ordering in a particle-hole channel.\cite{vafek13} 

Let us now consider the effect of a nonzero chemical potential. (We shall assume that $\mu \geq 0$ for concreteness; the results are the same for $\mu \leq 0$ by particle-hole symmetry.) At $T=0$ the chemical potential flows according to its engineering dimension as $\mu_\ell = \mu_0 e^{2\ell}$. The flow equations will not be affected by $\mu$ so long as $\mu_\ell \ll \Lambda^2 / 2m^*$. But as $\ell$ approaches $\ell_\mathrm{FS}$, which is defined such that $\mu_{\ell_\mathrm{FS}} = \Lambda^2 / 2m^*$, the flows will deviate from their $\mu=0$ behavior, allowing for three possible cases, shown in Figure \ref{fig:g_flows0}(e). (i) If $\mu_0$ is chosen to be very small, such that $\ell_\mathrm{FS} \gg \ell_*$, then the couplings will diverge before the chemical potential has any appreciable effect on the flows, and the system will again flow to a particle-hole ordered phase. (ii) If $\mu_0$ is chosen to be very large, such that $\ell_\mathrm{FS} \ll \ell_1$, then the UV cutoff reaches the chemical potential before an attractive coupling is generated. In this case there is no instability, and one obtains a Fermi liquid. (Due to the fact that some bare couplings $\tilde g_i (0)$ vanish for Hubbard interaction, it is not possible to have a Fermi liquid ground state in this special case.) Finally, (iii) if $\mu_0$ is chosen at some intermediate value with $\ell_\mathrm{FS}$ between $\ell_1$ and $\ell_*$, then the UV cutoff reaches the Fermi surface after an attractive coupling has been generated, but before the particle-hole instability at $\ell_*$. In this case there is a runaway flow of the most negative coupling only, and a superconducting phase is realized.

As noted previously\cite{vafek13}, the flow equations exhibit scaling behavior when $T = \mu = v_3 = 0$, in which case the coefficients $\mathcal{A}_{ijk}$ in \eqref{g_flow} become simple numbers, and one has solutions of the form 
\be
\label{eq:g_scaling}
g_i (\ell, \{g_j(0)\}) = g \Phi_i (g\ell, \{g_j(0)/g\}),
\ee
where $g = \sqrt{\sum_{i=1}^9 g_i^2(0)}$ is the overall magnitude of the bare coupling. This scaling behavior allows us to make two important statements about the flows. First, since all couplings are proportional to $g$, the magnitude of the couplings can be made arbitrarily small in the vicinity of $\ell_1$, where attraction is generated, by choosing $g$ to be appropriately small. Second, because of the argument $g\ell$ appearing in the scaling function $\Phi_i$, one can always satisfy $\mu_{\ell_1} \ll \Lambda^2/2m^* \ll \mu_{\ell_*}$, or equivalently $e^{-2\ell_*} \ll e^{-2\ell_\mathrm{FS}} \ll e^{-2\ell_1}$ by making $g$ sufficiently small, due to the fact that $\ell_1 , \ell_* \propto 1/g$. This means that it is always possible to have $\mu_{\ell_1}$ sufficiently small such that the flows are unaffected by chemical potential up to this point, but still have $\ell_\mathrm{FS} < \ell_*$, so that superconductivity is realized. Thus the scaling relation \eqref{eq:g_scaling} shows that the theory is controlled in the weak-coupling limit.

For larger values of chemical potential, as shown in Figure \ref{fig:g_flows0}(b,d), one finds that the above arguments are borne out, with some couplings turning attractive, as before, while others remain repulsive. As $\ell$ is increased further, however, all of the couplings saturate, except for the most attractive one, which flows to $-\infty$. In this case, the combination of attractive interaction in a pairing channel together with the kinematic constraints associated with the presence of a large Fermi surface lead to superconductivity, as one can verify by calculating pairing susceptibilities.

\section{Solution of flow equations at finite temperature}
\label{sec:temperature}
Let us continue by analyzing the solutions at nonzero temperature (but still with $v_3 =0$). Rather than starting from the $T=0$ limit discussed in Section \ref{sec:temperature} and introducing a small temperature, in which case the behavior is qualitatively similar to that already discussed, we instead take the approach of starting from high temperature and then decreasing $T$ until the couplings and susceptibilities start to become large, indicating the approach to a phase instability. In this way the bare temperature $T$ can be used as a control parameter, and the results obtained in this way can be seen as complementary to those obtained at $T=0$. In solving the flow equations, it is generically found that, for sufficiently high temperatures, some subset of the couplings $g_i (\ell)$ initially grow in magnitude before saturating to finite values as $\ell \to \infty$. As temperature is lowered, this saturation occurs at increasingly large values of $\ell$, with $g_i (\ell)$ increasing exponentially in magnitude over some range of $\ell$ before eventually saturating. At some critical temperature, which we denote as $T_{e(c)}$ for excitonic (superconducting) instabilities, the coupling magnitudes increase indefinitely without saturation as $\ell \to \infty$. 

\begin{figure}
\centering
\includegraphics[width=0.48\textwidth]{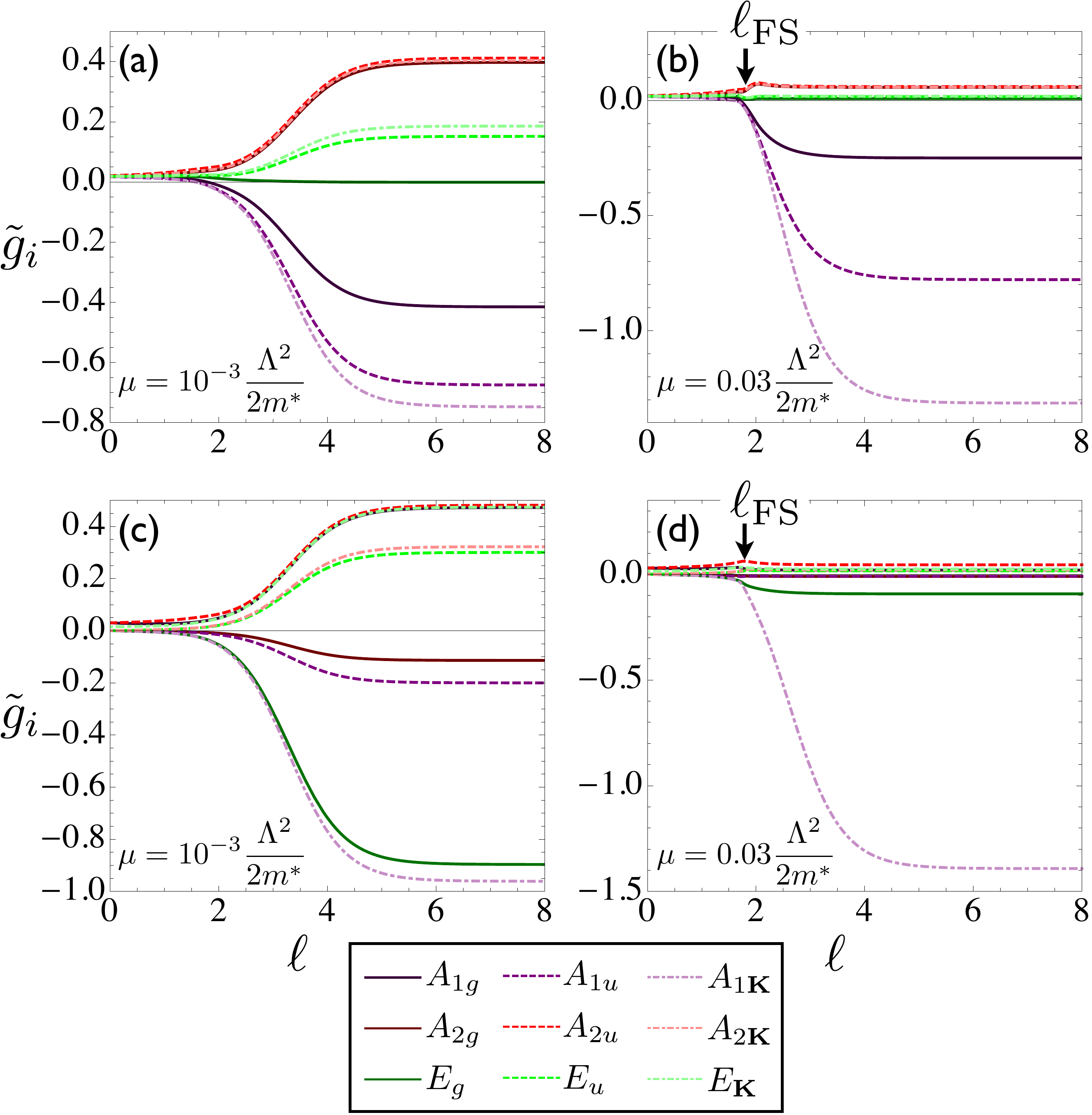}
\caption{RG flows of the couplings $\tilde g_i (\ell)$ for $T\gtrsim T_{e,c}$ and with $v_3 = 0$, for various values of chemical potential. Upper panels are calculated for near-forward scattering interaction, with bare couplings $g_{A_{1g}} = 0.15$, $g_{A_{2u}} = g_{E_\mathbf{K}} = 0.003$. Instabilities are to nematic (a) and pair density wave (b) phases. Lower panels are calculated for Hubbard interaction, with bare couplings $g_{A_{1g}} = g_{A_{2u}} = 0.06$, $ g_{E_\mathbf{K}} = 0.03$. Instabilities are to antiferromagnetic (c) and pair density wave (d) phases.
\label{fig:g_flows1}}
\end{figure}

The flows of the particle-particle couplings $\tilde g_i (\ell)$ for $T\gtrsim T_{c,e}$ are shown in Figure \ref{fig:g_flows1}.
In both cases shown in Figure \ref{fig:g_flows1}, one finds that---even when all bare couplings are repulsive---some attractive couplings are generated before the values of the couplings become large ($\sim \pm 1$). In the cases where particle-hole order is realized (Figure \ref{fig:g_flows1}(a,c)), several couplings approach $\pm \infty$ at $T=T_e$ as $\ell \to \infty$. For the cases where particle-particle order is realized (Figure \ref{fig:g_flows1}(b,d)), however, the flows are initially similar, but one finds that all couplings saturate upon integrating through the Fermi surface at $\ell = \ell_\mathrm{FS}$ except for the one corresponding to the superconducting order, for which $\tilde g_i \to -\infty$. Note that, unlike the $T=0$ case described in Section \ref{sec:mu}, there is no singularity upon integrating through the Fermi surface as long as $T>T_{e,c}$. 

With the above examples in mind, let us proceed to discuss the general aspects of the RG flows in the limit of large $\ell$. (The details of this analysis can be found in Appendices \ref{sec:bilayer_g_mu_flows} and \ref{sec:bilayer_susceptibilities}.) In analyzing the asymptotic flow equations, one finds two distinct regimes with very different behaviors. In the first, which occurs for sufficiently small values of the bare chemical potential, we find that $\mu_{\ell \to \infty} \sim e^{\alpha \ell}$, with $\alpha < 2$, and the diverging couplings blowing up as $g_i(\ell\to\infty) \sim e^{2\ell}$. Due to the fact that $\mu_\ell$ is less relevant than temperature in this case, the coefficients in the flow equations \eqref{mu_flow}, \eqref{g_flow}, and \eqref{vertex_flow} depend only on temperature as $\ell\to\infty$. Thus, while the presence of a chemical potential may affect nonuniversal properties such as the critical temperature, the critical exponents and possible phases that can be realized do not depend on $\mu$ in this regime. We thus recover the fixed ratios and associated phases, almost all of which correspond to particle-hole instabilities, from the half-filled case\cite{cvetkovic12} (see also Table \ref{table:matrices}).

In the second regime, which occurs for sufficiently large bare values of the chemical potential, we again find that $\mu_{\ell \to \infty} \sim e^{\alpha \ell}$, but now with $\alpha > 2$, while for the diverging couplings, $g_i(\ell\to\infty) \sim e^{\alpha \ell}$. In this case the chemical potential is {\em more} relevant than temperature, and the flow equation coefficients turn out to depend only on $\mu_\ell$. In this limit, only the particle-particle ladder diagrams from Figure \ref{RG} contribute to the flow equations for the couplings, and---in contrast to the particle-hole case---the critical exponents assume mean-field values, with $\eta_i^{pp} = 2$ in the divergent channel. The couplings then approach an entirely different set of fixed ratios, all of which correspond to superconducting instabilities. There are 9 of these fixed ratios, with one corresponding to each irreducible representation, as shown in Table \ref{table:matrices}. 

We find that generically, {\em i.e.}\ for any initial choice of repulsive interactions, there is a crossover between the two regimes described above, with particle-hole order giving way to an unconventional superconducting phase as the chemical potential is increased. From this analysis we see that such crossover behavior can be usefully described as a competition between temperature and chemical potential to be the most relevant parameter (in the RG sense), with the winner of this competition ultimately determining which type of phase is realized.

In order to ascertain phase instabilities, it is necessary to compute susceptibilities by analyzing the asymptotic behavior of the flow equations analytically at $T = T_c$. Although the couplings $g_i(\ell)$ diverge as $\ell\to\infty$, their ratios approach fixed finite values, with these fixed ratios ultimately determining the nature of the phase that is realized.\cite{cvetkovic12} The instabilities are determined by the anomalous critical exponents $\eta_i^{ph,pp}$, which are defined by the relations
\be
\frac{d \ln \Delta_i^{ph,pp}}{d\ell} & \stackrel{(\ell \to \infty)}{=} 
\begin{cases} 
2 + \eta_i^{ph,pp} ,  & \alpha < 2, \\
2 + \alpha \eta_i^{ph,pp}/2 ,  & \alpha > 2,
\end{cases}
\ee
where the explicit expressions for $\alpha$ and $\eta_i^{ph,pp}$ in terms of the fixed coupling ratios are provided in Appendices \ref{sec:bilayer_g_mu_flows} and \ref{sec:bilayer_susceptibilities}. Calculating the susceptibilities associated with the source terms \eqref{source_terms}, one finds that $\chi_i^{ph,pp} (T) \sim (T - T_c)^{1-\eta_i^{ph,pp}}$, so that there is an instability toward a particular phase when the associated anomalous critical exponent satisfies $\eta_i^{ph,pp} > 1$.

The phase diagrams obtained from the full numerical solution of the flow equations for various choices of initial parameters are shown in Figure \ref{fig:phases3d}.
In particular, for $v_3=0$ and near-forward scattering interaction, we find that the nematic phase, which breaks the 3-fold rotational symmetry of the lattice and transforms according to the $E_g$ representation, is favored for small $\mu$ in the near-forward scattering limit, both with and without trigonal warping. This is in agreement with some previous theoretical \cite{lemonik10,vafek10b,cvetkovic12} and experimental \cite{mayorov11} results. Upon increasing the chemical potential, this particle-hole order is suppressed in favor of a superconducting phase. Without trigonal warping, the superconducting phase is a pair density wave (PDW) state transforming according to the $A_{1\mathbf{K}}$ representation. In this state, the electrons from {\em within} a single Fermi pocket pair with one another, so that the Cooper pairs carry nonzero total momentum $\mathbf{q} = 2\mathbf{K}$. The realization of such PDW superconductivity, originally proposed theoretically by Fulde, Ferrel, Larkin, and Ovchinnikov \cite{fulde64,larkin65}, has been a longstanding experimental challenge. As we see in Figure \ref{fig:phases3d} and shall describe in Section \ref{sec:trig}, however,  the inclusion of trigonal warping tends to suppress the PDW in favor of other superconducting phases. For the case of bare Hubbard interaction, shown in Figure \ref{fig:phases3d}(b), the particle-hole instability is toward a layer antiferromagnetic phase, in agreement with some previous theoretical \cite{castro_neto09,vafek10,kharitonov12} and experimental \cite{velasco12,bao12} results. Upon doping, one obtains again the PDW superconducting phase. The magnitude of the bare coupling in each case is chosen to yield a maximum critical temperature $T_e \sim 0.01 \Lambda^2 / 2m^*$ for the excitonic instability. This corresponds roughly to the energy scales ($\sim 1$ meV) at which signatures of ordering have been observed experimentally.

We emphasize that the RG approach used here is only able to determine phase instabilities when approaching the transition from temperatures $T>T_c$. The phase diagrams shown in Figure \ref{fig:phases3d} are thus unable to address the behavior for $T\ll T_c$, {\em e.g.}\ the possibility of other orders being induced or of coexistence between excitonic and superconducting orders. Indeed, from the $T=0$ flows shown in Figure \ref{fig:g_flows0}(b), one expects a PDW phase at doping $\mu = 0.013 \Lambda^2 / 2m^*$, which is directly beneath the nematic (rather than the PDW) instability shown in Figure \ref{fig:phases3d}. These results are not necessarily inconsistent, however, as the PDW may extend to smaller values of $\mu$ at $T \ll T_{e,c}$. Furthermore, the temperatures $T_e$ and $T_c$ should be understood as {\em approximate} ordering temperatures, due to the fact that the weak-coupling approach breaks down very close to these temperatures, where the renormalized couplings and susceptibilities become large. In addition, fluctuations will suppress true long-range order of a phase breaking continuous symmetry in a two-dimensional system, so in such cases it is more appropriate to interpret $T_e$ and $T_c$ as characteristic temperature scales where ordering behavior ({\em e.g.}\ power-law correlations, corresponding to quasi-long range order) sets in.

\section{Solution of the flow equations with trigonal warping}
\label{sec:trig}
\begin{figure}
\centering
\includegraphics[width=0.35\textwidth]{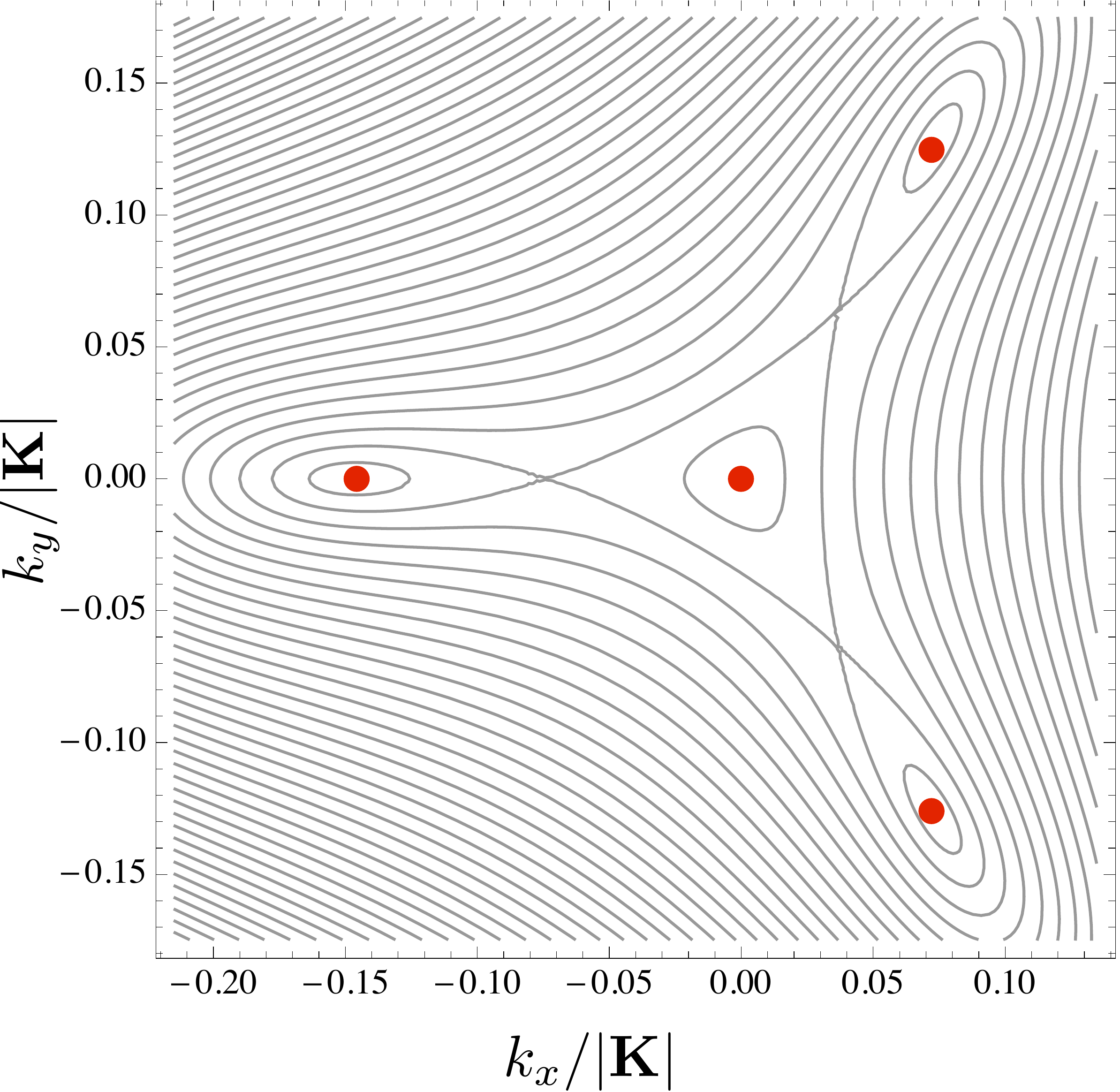}
\caption{Contours of constant energy for electron dispersion in the presence of trigonal warping, with $v_3 = 0.178 \Lambda / 2m^*$, and contours separated by $\Delta E = 8\times10^{-3} \Lambda^2 / 2m^*$. The point $\bk=(0,0)$ corresponds to the Brillouin zone corner at $+\mathbf{K} = (\frac{4\pi}{3\sqrt{3}a},0)$. The red points show the locations of the four Dirac points, with linear dispersion about each of these points.
\label{fig:trig}}
\end{figure}

Let us now describe the effects of trigonal warping, which is characterized by the velocity $v_3$ appearing in \eqref{H0} and arises from further-neighbor interlayer hopping ($v_3 \sim \gamma_3$ in Figure \ref{bilayer_dispersion}). The main effect is to modify the low-energy dispersion from parabolically touching bands to four miniature Dirac cones near each Brillouin zone corner $\pm\mathbf{K}$, as shown in Figure \ref{fig:trig}. 
Such a distortion should generically be present in the noninteracting band structure of realistic materials such as bilayer graphene, and including it in our calculations allows us to better understand the robustness of our mechanism for unconventional superconductivity more generally away from special fine-tuned cases.

\begin{figure}
\centering
\includegraphics[width=0.48\textwidth]{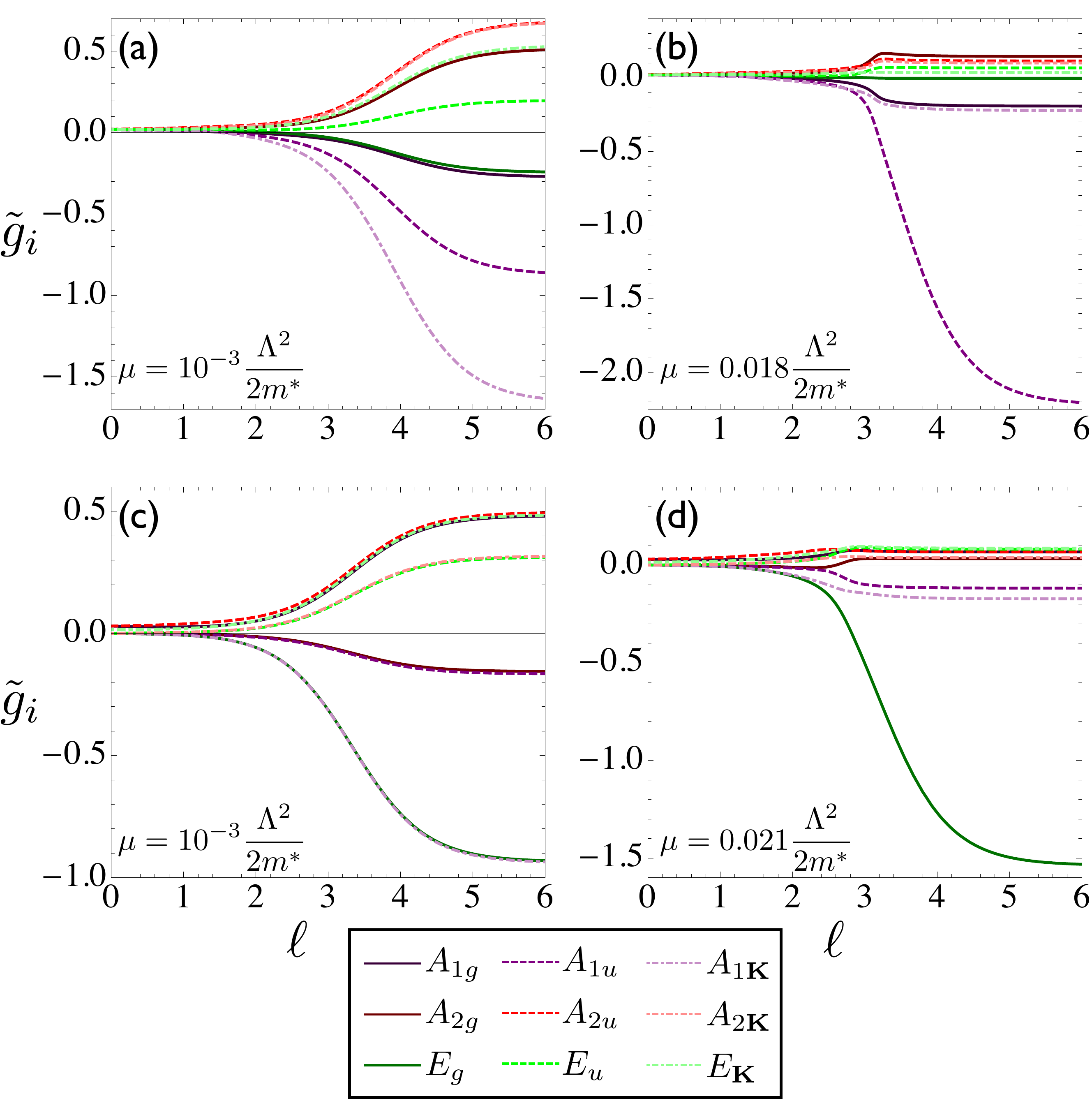}
\caption{RG flows of the couplings $\tilde g_i (\ell)$ for $T\gtrsim T_{e,c}$ and with $v_3 = 0.178 \Lambda/2m^*$, for various values of chemical potential. Upper panels are calculated for near-forward scattering interaction, with bare couplings $g_{A_{1g}} = 0.15$, $g_{A_{2u}} = g_{E_\mathbf{K}} = 0.003$. Instabilities are to nematic (a) and $f$-wave superconducting (b) phases. Lower panels are calculated for Hubbard interaction, with bare couplings $g_{A_{1g}} = g_{A_{2u}} = 0.06$, $ g_{E_\mathbf{K}} = 0.03$. Instabilities are to antiferromagnetic (c) and $d$-wave superconducting (d) phases.
\label{fig:g_flows2}}
\end{figure}

The solutions to the flow equations for the couplings $\tilde g_i(\ell)$ with finite trigonal warping are shown in Figure \ref{fig:g_flows2}.
The behavior is qualitatively similar to that for the $v_3=0$ case shown in Figure \ref{fig:g_flows1}, except that the most negative couplings near the superconducting instability correspond to $f$-wave and $d$-wave channels, rather than PDW as before. It is natural that the PDW should be suppressed, due to the fact that---unlike the $f$- and $d$-wave states---it is sensitive to the $\bk \to -\bk$ symmetry within a pocket, which is destroyed by trigonal warping. This does not occur immediately upon turning on $v_3$, however. For the choices of couplings illustrated in Figure \ref{fig:g_flows2}, the critical value separating the PDW and $f$-wave phases for near-forward scattering is $v_3 \approx 0.006\Lambda / 2m^*$, while the critical value separating the PDW and $d$-wave phases for Hubbard interaction is $v_3 \approx 0.034\Lambda / 2m^*$. In both cases, the critical value is substantially less than the experimentally estimated\cite{mayorov11} value of $v_3 \approx 0.178\Lambda / 2m^*$ for bilayer graphene, suggesting that the $f$- and $d$-wave phases are more likely to be realized experimentally. 

The phase diagrams obtained from solving the finite-temperature flows with finite $v_3$ are shown in Figure \ref{fig:phases3d}.
Due to the fact that $v_3$ is less relevant than temperature under RG flow, the $\ell\to\infty$ analysis of Section \ref{sec:temperature} remains valid for $v_3 \neq 0$. As in the case with $v_3 = 0$, the ordering temperatures for the superconducting phases are smaller than, but of the same order as, those for excitonic phases. The $f$-wave superconducting phase, which is realized for long-ranged interactions, transforms according to the $A_{1u}$ representation and features a sign change of the order parameter between pockets at $\pm \mathbf{K}$. The $d$-wave phase, which is realized for short-ranged interaction, transforms according to the two-dimensional $E_g$ representation. Due to the presence of two complex order parameter components, the latter phase has a richer phenomenology, which shall be studied in detail in Section \ref{sec:SC}.

Figure \ref{fig:phases3d} clearly shows that both excitonic and superconducting critical temperatures are suppressed due to trigonal warping. Indeed, due to the fact that contact interactions are irrelevant for linearly dispersing fermions in two dimensions, there are no longer instabilities to ordered phases for arbitrarily small $g_i$ at fixed $v_3\neq 0$.
It is instructive to compare the ordering temperatures with the van Hove energy $E_\mathrm{vH} = 2m^* v_3^2$, which is the characteristic energy scale associated with trigonal warping. One finds that the ordered phases may persist for $E_\mathrm{vH} \gg T_{e,c}$ with the quantities differing by an order of magnitude in the case of Hubbard interaction. In particular, for $\mu < E_\mathrm{vH}$ the Fermi surface is made up of four disconnected patches near each $\pm\mathbf{K}$. In this regime the low-energy behavior is dominated by the linear dispersion for $E<E_\mathrm{vH}$ rather than the quasi-quadratic dispersion at higher energies (although the latter remains necessary for generating attractive interactions). 

\section{Nature of the superconducting phase}
\label{sec:SC}
In Sections \ref{sec:temperature} and \ref{sec:trig} we found two cases in which the leading superconducting instability was to a phase transforming as a two-dimensional representation with respect to the lattice space group. For small values of $v_3$, the leading superconducting instability was found to be the PDW phase, transforming according to the $A_{1\mathbf{K}}$ representation. For larger values of $v_3$ and short-ranged interaction, the $d$-wave superconducting phase, transforming according to the $E_g$ representation, was found to be the leading instability. In principle, any linear combination of order parameter components within each of these cases might be realized, and our symmetry-based RG approach is unable to distinguish between them. In order to complete the analysis, then, one must supplement the RG analysis with another approach.

In this section we employ two such approaches. The first is a free energy expansion near the critical temperature, where a Landau free energy is derived from the microscopic theory by decoupling the interaction term in the ordering channel via Hubbard-Stratonovich transformation. Of course, this is not done with the bare couplings, which are repulsive, but rather with the attractive couplings obtained after running the RG to some intermediate scale $\ell_\mathrm{stop} \lesssim \ell_\mathrm{FS}$. The free energy coefficients are determined from the microscopic theory with these couplings, and these coefficients determine the nature of the ordered phase. This approach is justified by the mean-field nature of the transition to the superconducting phase, which is apparent from the mean-field values of the critical exponents $\eta_i$, as well as from the fact that only a single coupling $\tilde g_i$ flows to large values under RG near the transition. (We point out that this mean-field decoupling would {\em not} be justified in treating the excitonic instabilities, which feature non-mean field critical exponents and multiple diverging couplings.) This approach is employed separately for the $d$-wave and PDW phases in the following two subsections. In the $d$-wave case we find that the chiral, time-reversal symmetry breaking combination of order parameters, often denoted as ``$d+id$,'' is realized. In the PDW phase, on the other hand, we find that the non-chiral state, in which the pairing amplitude---but not the phase---modulates in space, is preferred.\cite{larkin65} 

The second approach used to study the superconducting phase is a self-consistent mean-field theory. This approach is valid within the ordered phase at low temperatures, thus complementing the  free energy expansion at $T\sim T_c$. This approach has the advantage that it allows one to address the competition between multiple types of order in cases where it is not unambiguously resolved from the RG calculation, for example when multiple couplings reach values $\sim \pm O(1)$ before some eventually saturate [{\em cf.} Figure \ref{fig:g_flows1}(b)]. In such cases it makes sense to supplement the RG results with mean-field calculations, which can be carried out after running the RG up to some intermediate scale $\ell = \ell_\mathrm{stop} \lesssim \ell_\mathrm{FS}$. In Section \ref{sec:mean_field} we use this method to address the competition between the three most likely superconducting instabilities: $f$-wave ($A_{1u}$), $d$-wave ($E_g$), and PDW ($A_{1\mathbf{K}}$). We also find agreement with the results from the free energy expansion in cases where the latter two superconducting orders are realized.

\subsection{Free energy expansion for $d$SC phase}
In Section \ref{sec:trig} it was determined that, in the presence of trigonal warping, a repulsive Hubbard interaction leads to a $d$-wave superconducting instability for sufficiently large chemical potential. However, the RG approach used thus far is unable to determine which linear combination of the two components ($d_{x^2 - y^2}, d_{xy}$) belonging to the $E_g$ representation will be realized. An intriguing possibility is that the order parameter components might coexist, with a relative phase between them (denoted as $d_{x^2 - y^2} + i d_{xy}$, or simply $d+id$), thereby breaking time-reversal symmetry. Such chiral phases have been proposed in a variety of condensed matter systems, with the hope of providing a solid-state analogue to the well-established\cite{wheatley75} $p+ip$ superfluidity in the A phase of $^3$He. Such a phase has likely already been observed\cite{maeno12} in Sr$_2$RuO$_4$. A chiral $d$-wave phase was first proposed in the context of high-temperature cuprate superconductors\cite{laughlin98}, and more recently there have been theoretical proposals of chiral $s$-wave phases in iron-based superconductors\cite{stanev10,maiti13}. 
Based on perturbative RG calculations, it was recently proposed that a time-reversal symmetry breaking combination of the two components ought to be realized in single-layer graphene doped to the van Hove point.\cite{nandkishore12} The possibility of $d+id$ superconductivity on the honeycomb bilayer has also been suggested recently \cite{vucicevic12}, although the strong-coupling mean-field theory used in that study did not account for the origin of the effective attractive interaction, nor did it address the competition of $d$-wave superconductivity with other ordered phases. The chiral $d$-wave phase has a rich phenomenology, and may feature spontaneous edge currents, as well as spin Hall and thermal Hall effects.\cite{volovik97,fogelstrom97,laughlin98,senthil99,horovitz03,black-schaffer12} In this section we address the question of whether chiral superconductivity can arise from repulsive interactions on the honeycomb bilayer via our weak-coupling RG analysis.

In the previous section it was shown that, if the couplings are flowing toward fixed ratios corresponding to a superconducting phase, then only one of the particle-particle couplings $\tilde g_i$ becomes large [See Figure \ref{fig:g_flows1}(b,d)]. In this case one is justified near $T_c$ in considering only fluctuations in the corresponding particle-particle channel. This allows for the interaction to be decoupled via a Hubbard-Stratonovich transformation, which in this case leads to the effective action
\ba
\label{S_Delta_psi}
S_{\Delta, \psi} = &\int d\tau \int d^2 x  \bigg\{ 
	\frac{1}{4 \tilde g_{E_g}} \left( |\Delta_{d1}|^2 + |\Delta_{d2} |^2 \right) \\
	&+ 2\big[ ( \Delta_{d1}^* - i\Delta_{d2}^* ) 
	\psi_{\mathbf{K}2\alpha} (\sigma_2)_{\alpha\beta} \psi_{-\mathbf{K}1\beta} \\
	&+ ( \Delta_{d1}^* + i\Delta_{d2}^* ) 
	\psi_{\mathbf{K}1\alpha} (\sigma_2)_{\alpha\beta} \psi_{-\mathbf{K}2\beta} + c.c. \big] \bigg\},
\ea
where $\Delta_{d1,d2}$ are the order parameters having $d_{x^2-y^2}$ and $d_{xy}$ symmetry, respectively, and $\alpha,\beta$ are spin indices. (Note that a factor of $i$ has been included in the definition of $\Delta_{d2}$, so that the time-reversal symmetry preserving state corresponds to $\Delta_{d1,d2}$ having the same complex phase.) Using standard methods, the fermionic degrees of freedom can be integrated out, leading to an effective action $S_\Delta$ for the superconducting fields $\Delta_{d1,d2}$. 

We next rewrite the Hubbard-Stratonovich action \eqref{S_Delta_psi} in Nambu spinor notation:
\ba
\label{S_Nambu}
S_{\Delta, \psi} &= T \sum_n \int \frac{d^2 k}{(2\pi )^2} 
	\Psi^\dagger_{n,k} \hat{\mathcal{G}}_\Delta^{-1} (i \omega_n, k) \Psi_{n,k} \\
	&+ \frac{1}{4\tilde g_{E_g}} \int d\tau \int d^2x
	 \left( |\Delta_{d1}|^2 + |\Delta_{d2} |^2 \right),
\ea
with
\be
\Psi_{n,\bk} = \left( \begin{matrix} \psi_{n,\bk,\uparrow} \\ \psi^*_{-n,-\bk,\downarrow}
	\end{matrix} \right).
\ee
The $8 \times 8$ Green function matrix is given by
\be
\label{G_Delta}
\hat{\mathcal{G}}_\Delta^{-1} (i \omega_n, k) = \hat{\mathcal{G}}_0^{-1} (i \omega_n, k) 
	+ \hat{\Delta}_d (i \omega_n, k),
\ee
where
\ba
\hat{\mathcal{G}}_0^{-1} (i \omega_n, k) &= -i \omega_n 1_{8} - \mu \rho_3 11 
	+ \frac{k_x^2 - k_y^2}{2 m^*} \rho_3 1 \sigma_1 \\
	&+ \frac{k_x k_y}{m^*} 1 \tau_3 \sigma_2
	+ v_3 k_x 1 \tau_3 \sigma_1 - v_3 k_y \rho_3 1 \sigma_2
\ea
is the bare Green function for fermions, and
\ba
\label{eq:Delta_d}
\hat{\Delta}_d =& \mathrm{Re} \Delta_{d1} \rho_1 \tau_1 \sigma_1 
	- \mathrm{Im} \Delta_{d1} \rho_2 \tau_1 \sigma_1 \\
	&+\mathrm{Im} \Delta_{d2} \rho_1 \tau_2 \sigma_2 
	+ \mathrm{Re} \Delta_{d2} \rho_2 \tau_2 \sigma_2 
\ea
The $2 \times 2$ matrices $\rho_i , \tau_i$, and $\sigma_i,$ appearing in these equations are Pauli matrices operating in Nambu, valley, and layer spaces, respectively.

Integrating out the fermions from the action \eqref{S_Nambu} yields the following effective action for the superconducting fields:
\ba
\label{S_expand}
S_\Delta =&  \frac{1}{4\tilde g_{E_g}} 
	\int d\tau \int d^2x \left( |\Delta_{d1}|^2 + |\Delta_{d2} |^2 \right) \\
	&\quad + \frac{1}{2} \mathrm{Tr} \left( \hat{\mathcal{G}}_0 \hat{\Delta_d} \right)^2 
	+ \frac{1}{4} \mathrm{Tr} \left( \hat{\mathcal{G}}_0 \hat{\Delta_d} \right)^4,	
\ea
where higher-order terms have been ignored. The traces in \eqref{S_expand} are over matrix indices, as well as frequency and momentum. The traces are most conveniently performed by letting
\be
\hat{\mathcal{G}}_0^{-1} (i \omega_n, k) = \left(
   \begin{matrix} 
      \hat{G}_+^{-1} (i \omega_n , k) & 0 \\
      0 & \hat{G}_-^{-1} (i \omega_n , k) \\
   \end{matrix} \right)
\ee
and
\be
\label{Delta8x8}
\hat{\Delta_d} = \left(
   \begin{matrix} 
      0 & \hat{\Delta}_{d1} - i \hat{\Delta}_{d2} \\
       \hat{\Delta}_{d1}^\dagger + i\hat{\Delta}_{d2}^\dagger  & 0 \\
   \end{matrix} \right).
\ee
In terms of these new matrices, the quadratic part of the action \eqref{S_expand} becomes
\ba
S_\Delta^{(2)} =& \int d\tau \int d^2x \left[
	\frac{1}{4\tilde g_{E_g}} \left( |\Delta_{d1}|^2 + |\Delta_{d2} |^2 \right) \right] \\
	&+ \mathrm{Tr} \left[ \hat{G}_+ (\hat{\Delta}_{d1} - i\hat{\Delta}_{d2}) 
	\hat{G}_-  ( \hat{\Delta}_{d1} - i\hat{\Delta}_{d2})^\dagger \right].
\ea
The trace in this expression can be evaluated using standard methods. Assuming that the $E_g$ coupling is the most negative, as found in our RG solutions, one finds a sign-changing term $\sim (T - T_c)(|\Delta_{d1}|^2 + |\Delta_{d2}|^2)$, so that there is a mean-field transition into the $d$-wave superconducting phase below temperature $T_c$, which is determined by the following condition:
\ba
\frac{1}{4\tilde g_{E_g}} = -\int \frac{d^2 k}{(2\pi)^2} \bigg[ &
	 \frac{1}{\xi^+_\bk + \mu} \tanh \left( \frac{\xi^+_\bk + \mu}{2T_c} \right) \\
	 &+ \frac{1}{\xi^+_\bk - \mu} \tanh \left( \frac{\xi^+_\bk - \mu}{2T_c} \right) \bigg].
\ea
Here we have defined
\be
\label{eq:1015l}
\xi_\bk^\pm = \sqrt{\varepsilon_\mathbf{k}^2 
	+ v_3^2 k^2 \pm 2 v_3 k \varepsilon_\mathbf{k} \cos 3\theta_\mathbf{k}}.
\ee

The nature of the superconducting phase is determined by the fourth-order term in \eqref{S_expand}:
\ba
\label{S_Delta4}
S_\Delta^{(4)} &= \mathrm{Tr} \bigg[ \hat{G}_+ (\hat{\Delta}_{d1} + i \hat{\Delta}_{d2}) 
	\hat{G}_-  (\hat{\Delta}_{d1} + i \hat{\Delta}_{d2})^\dagger \bigg]^2 \\
	&= \int d^2 x \int d\tau \bigg[ 
	\frac{\beta_d}{2} \left( |\Delta_{d1} (\tau, x)|^2 + |\Delta_{d2} (\tau, x)|^2 \right)^2 \quad \\
	&\quad\quad\quad\quad + \gamma_d |\Delta_{d1}^2 (\tau, x) + \Delta_{d2}^2 (\tau, x)|^2 \bigg].
\ea
Equation \eqref{S_Delta4} is in fact the most general possible form of a quartic contribution to the free energy that is invariant under the symmetry of the honeycomb lattice.\cite{mineev98,SCfootnote} The first coefficient $\beta_d$ is positive, and its precise value shall not be of concern here. The second coefficient in \eqref{S_Delta4} is given by
\begin{widetext}
\be
\label{gamma1}
\gamma_d = T\sum_n \int \frac{d^2 k}{(2\pi)^2} 
	\frac{(\xi^+_\bk)^4 + 2(\xi^+_\bk)^2 (\mu^2-\omega_n^2) - (\mu^2 + \omega_n^2)^2 
	-2 [ \varepsilon_\bk^2 \cos 4\theta_\bk + (v_3 k)^2 \cos 2\theta_\bk 
	- 2v_3 k \varepsilon_\bk \cos 3\theta_\bk ]^2}{[(i\omega_n + \mu)^2 - (\xi^+_\bk)^2]^2[(i\omega_n - \mu)^2 - (\xi^+_\bk)^2]^2} ,
\ee
\end{widetext}
where we have set external momenta and frequencies to zero when performing traces over the Green's functions. For $T < T_c$, the nature of the superconducting phase depends on the sign of $\gamma_d$. Below we consider the two possible cases in turn.

For $\gamma_d < 0$, the free energy is minimized by maximizing the amplitude of the last term in \eqref{S_Delta4}. This clearly occurs when there is no relative phase difference between $\Delta_{d1}$ and $\Delta_{d2}$. In this case, assuming $\Delta_i (\tau, x) = const.$, the free energy density $f$ is given by 
\ba
f =& \alpha \left( |\Delta_{d1}|^2 + |\Delta_{d2}|^2 \right) \\
	&+ \left( \frac{\beta_d}{2} + \gamma_d \right) \left( |\Delta_{d1}|^2 + |\Delta_{d2}|^2 \right)^2.
\ea
Clearly the phase is only stable when $\beta_d / 2 + \gamma_d > 0$. If this is the case, the free energy is minimized for $\Delta_{d1} = \Delta_{d0} \cos \theta$ and $\Delta_{d2} = \Delta_{d0} \sin \theta$, where $\Delta_{d0} = \sqrt{-\alpha / (\beta_d + 2 \gamma_d)}$, and $\theta$ can take any value. Because there is no phase difference between the two order parameter components, this phase is non-chiral and preserves time-reversal symmetry. 

For $\gamma_d > 0$, on the other hand, the last term in \eqref{S_Delta4} is clearly minimized when $\Delta_{d2} = e^{\pm i \pi / 2} \Delta_{d1}$. 
Because the two order parameter components coexist with a nontrivial relative phase between them, time-reversal symmetry is broken in this case. This is the chiral, time-reversal symmetry breaking $d+id$ phase.

Having identified these two possibilities, the next step is to determine the sign of $\gamma_d$ from Equation \eqref{gamma1}. The equation takes on a slightly more tractable form in the limit $v_3 \to 0$:
\ba
\gamma_d \stackrel{(v_3=0)}{=}& T \sum_n \int \frac{d^2 k}{(2\pi)^2} \\
	& \quad \times \frac{2\varepsilon_\bk^2 (\mu^2-\omega_n^2) - (\mu^2 + \omega_n^2)^2}{[(i\omega_n + \mu)^2 - \varepsilon_\bk^2]^2[(i\omega_n - \mu)^2 - \varepsilon_\bk^2]^2}.
\ea
In the weak-coupling limit, this integral is dominated by the infrared singularity at $\varepsilon_\bk \sim \mu$, $\omega_n \sim T_c \ll \mu$ (recall that the Ginzburg-Landau expansion we are performing is valid only for $T\sim T_c$). In this case one sees that the above integral is positive, as can be verified numerically, and so the chiral phase is indeed stable.
\begin{figure}
\centering
\includegraphics[width=0.47\textwidth]{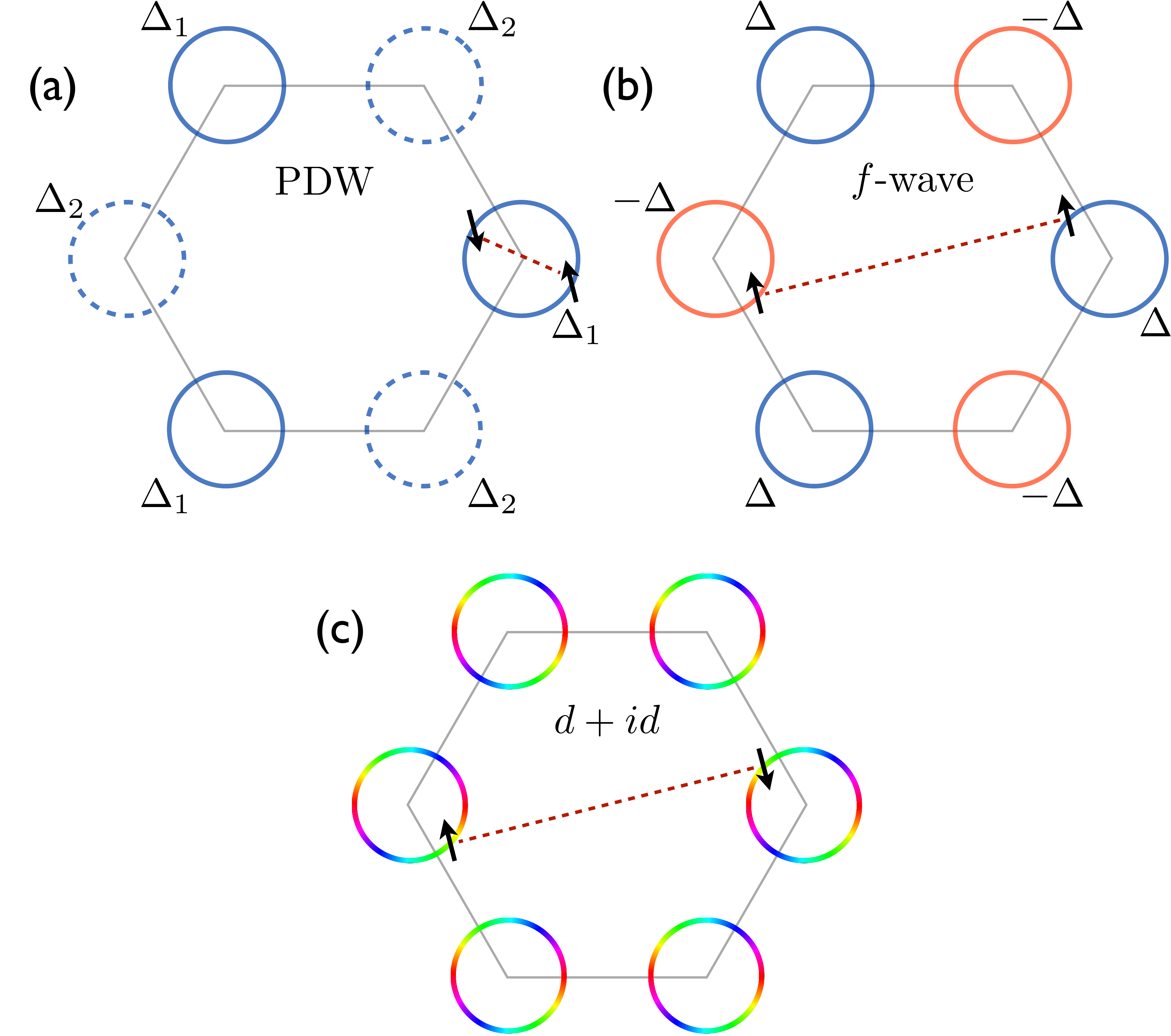}
\caption{Superconducting states in momentum space. (a) In the PDW ($A_{1\mathbf{K}}$) phase, which is realized for small $v_3$, Cooper pairs are formed by electrons within the same pocket and so have nonzero total momentum. The order parameters $\Delta_{1,2}$ have the same complex phase but arbitrary relative amplitude. (b) In the $f$-wave ($A_{1u}$) superconducting phase, which is realized for longer-ranged interactions, the pairing is between pockets, and the order parameter has uniform amplitude and opposite sign on the Fermi pockets at $\pm \mathbf{K}$. (c) In the chiral $d$-wave ($E_g$) state, the pairing is again between pockets. Each pocket is fully gapped, and the complex phase of the order parameter (represented by color) winds by $\pm 4\pi$ around each pocket.
\label{fig:SC}}
\end{figure}
As shown in Figure \ref{fig:SC}(c), this state is fully gapped and features a winding of the complex phase by $\pm 4\pi$ as one circles around a Fermi pocket.

\subsection{Free energy expansion for PDW phase}
For small values of $v_3$, it was found in Sections \ref{sec:mu} and \ref{sec:temperature} that the PDW superconducting phase, which belongs to the $A_{1\mathbf{K}}$ representation, is the leading instability. As in the $d$-wave case, this representation is two-dimensional, and the RG procedure alone does not determine which linear combination of the two order parameter components is selected. In order to determine this, we again derive an effective Landau free energy as in the previous section, but now with the order parameter  
\ba
\label{eq:1015c}
\hat{\Delta}_\mathbf{K} =& \mathrm{Re} \Delta_{\mathbf{K}1} \rho_1 1 \sigma_1 
	- \mathrm{Im} \Delta_{\mathbf{K}1} \rho_2 1 \sigma_1 \\
	&+\mathrm{Im} \Delta_{\mathbf{K}2} \rho_1 \tau_3 \sigma_1 
	+ \mathrm{Re} \Delta_{\mathbf{K}2} \rho_2 \tau_3 \sigma_1 .
\ea
(As in the $d$-wave case, a factor of $i$ has been included in the definition of $\Delta_{\mathbf{K}2}$, such that the combination preserving time-reversal symmetry corresponds to $\Delta_{\mathbf{K}1}$ and $\Delta_{\mathbf{K}2}$ having the same phase.)

As in the previous subsection, the nature of the PDW phase is determined by the fourth-order term in the Ginzburg-Landau expansion for the effective action:
\be
\label{eq:1015i}
S_\Delta^{(4)} = \frac{1}{4} \rm{Tr} \left( \hat{\mathcal{G}}_0 \hat{\Delta}_\mathbf{K} \right)^4 .
\ee
Evaluating the trace and simplifying gives the following 4th order term for the free energy density:
\be
\label{eq:1015j}
f^{(4)} = \frac{\beta_\mathbf{K}}{2} ( |\Delta_{\mathbf{K}1}|^2 + |\Delta_{\mathbf{K}2}|^2)^2
	+ \gamma_\mathbf{K} |\Delta_{\mathbf{K}1}^2 + \Delta_{\mathbf{K}2}^2|^2,
\ee
where, as usual, the frequency or momentum dependence of $\Delta_{\mathbf{K}1,\mathbf{K}2}$ have been set to zero. The coefficients are given by
\begin{widetext}
\be
\label{eq:1015k}
\beta_\mathbf{K} = 8T \sum_n \int \frac{d^2 k}{(2\pi)^2} 
	\frac{[(i\omega_n)^2 - \mu^2]^2 - 16(\varepsilon_\bk v_3 k \sin 3\theta_\bk)^2
	+ 2\mu^2 [3\varepsilon_\bk^2 - (v_3 k)^2] 
	+ 2(i\omega_n)^2[3 (v_3 k)^2 - \varepsilon_\bk^2]}{[(i\omega_n + \mu)^2 + (\xi_\bk^+)^2]^2 [(i\omega_n - \mu)^2 + (\xi_\bk^-)^2]^2} 
\ee
\end{widetext}
and $\gamma_\mathbf{K} = -\beta_\mathbf{K}/4$. In contrast to the $d$-wave case, here we find that $\gamma<0$, implying that the non-chiral PDW phase is stable. In this state, the order parameter amplitude varies spatially as either $\Delta_{\mathbf{K}1} \sim \cos (2\mathbf{K}\cdot\mathbf{x} )$ or $\Delta_{\mathbf{K}2} \sim \sin (2\mathbf{K}\cdot\mathbf{x} )$, or some linear combination of the two, with only the overall amplitude $|\Delta_{\mathbf{K}1}|^2 + |\Delta_{\mathbf{K}2}|^2$ fixed by the minimization of the free energy \eqref{eq:1015j}, and the two components having the same complex phase. It may be possible for the amplitude modulation at wavevector $\mathbf{K}$ to be detected experimentally using probes such as scanning tunneling microscopy or transmission electron microscopy, although a detailed study of the phenomenology of this phase is left for future work. 

\subsection{Self-consistent mean-field solution}
\label{sec:mean_field}
In solving the RG flow equations, in some cases we find that, although a single coupling ultimately becomes the most negative and determines the superconducting state, one or more other couplings may grow together with it, and these other couplings may not saturate until fairly late in the RG flow. This is the case, for example, for the flows shown in Figure \ref{fig:g_flows0}(b), where the coupling for the $f$-wave channel is nearly degenerate with the PDW coupling over most of the flow, and also in Figure \ref{fig:g_flows0}(d), where the $f$-wave and $d$-wave couplings are nearly degenerate. In some cases where the bare couplings are near an unstable fixed ratio, the flows may follow this unstable trajectory until the couplings approach values $\sim \pm 1$, in which case the weak-coupling RG approach begins to break down. This occurs for example very close to the pure forward-scattering limit, which follows an unstable flow in which the couplings and susceptibilities in the $A_{1g}$, $A_{1u}$, and $A_{1\mathbf{K}}$ are degenerate.\cite{vafek13} In such cases it is useful to supplement the RG approach with a self-consistent mean-field treatment. 
In this hybrid approach, we first run the RG up to $\ell = \ell_\mathrm{stop}$, which is chosen to be past the point where attraction is generated, but before the couplings become large. The values of the couplings and other parameters at $\ell_\mathrm{stop}$ are then used as inputs in a self-consistent mean-field calculation for the superconducting order parameters. 

As before, the Bogoliubov-de Gennes Green function is given by
\be
\label{eq:1104a}
\hat{\mathcal{G}}^{-1}_\Delta (i \omega_n, k) = \hat{\mathcal{G}}_0^{-1} (i \omega_n, k) + \hat{\Delta},
\ee
where now the order parameter matrix includes all three of the most likely superconducting orders:
\be
\hat \Delta = \hat\Delta_d + \hat\Delta_\mathbf{K} + \hat\Delta_f,
\ee
with $\hat\Delta_d$ and $\hat\Delta_\mathbf{K}$ from \eqref{eq:Delta_d} and \eqref{eq:1015c}, and
\be
\hat\Delta_f = \mathrm{Re} \Delta_{A1u} \rho_1 \tau_2 1
	+ \mathrm{Im} \Delta_{A1u} \rho_2 \tau_2 1,
\ee 
corresponding to $f$-wave ($A_{1u}$) superconducting order. It is straightforward to include other superconducting order parameters as well, but they will vanish in the mean-field solution unless their corresponding couplings are negative and comparable to $\tilde g_{E_g}$, $\tilde g_{A_{1\mathbf{K}}}$, and $\tilde g_{A_{1u}}$.
From \eqref{eq:1104a} one obtains the following self-consistent mean-field equations:
\be
\label{eq:1104d}
\Delta_i = \frac{\tilde g_i}{L^2} \sum_{|\bk|<\Lambda} \mathrm{Tr} 
	(M_i \langle \Psi_\bk \Psi_\bk^T \rangle),
\ee
where $L^2$ is the number of states within the momentum cutoff, and $M_i$ is the $8\times 8$ Nambu matrix corresponding to a particular order parameter component. For example,
\be
\label{eq:1104e}
\mathrm{Re}\Delta_{d1} = \frac{\tilde g_{E_g}}{L^2} \sum_{|\bk|<\Lambda} \mathrm{Tr} 
	(\rho_1\tau_1 \sigma_1 \langle \Psi_\bk \Psi_\bk^T \rangle).
\ee
The matrix of expectation values $\langle \Psi_\bk \Psi_\bk^T \rangle$ is computed by diagonalizing the Bogoliubov-de Gennes Hamiltonian at each point in momentum space. 

\begin{figure}
\centering
\includegraphics[width=0.45\textwidth]{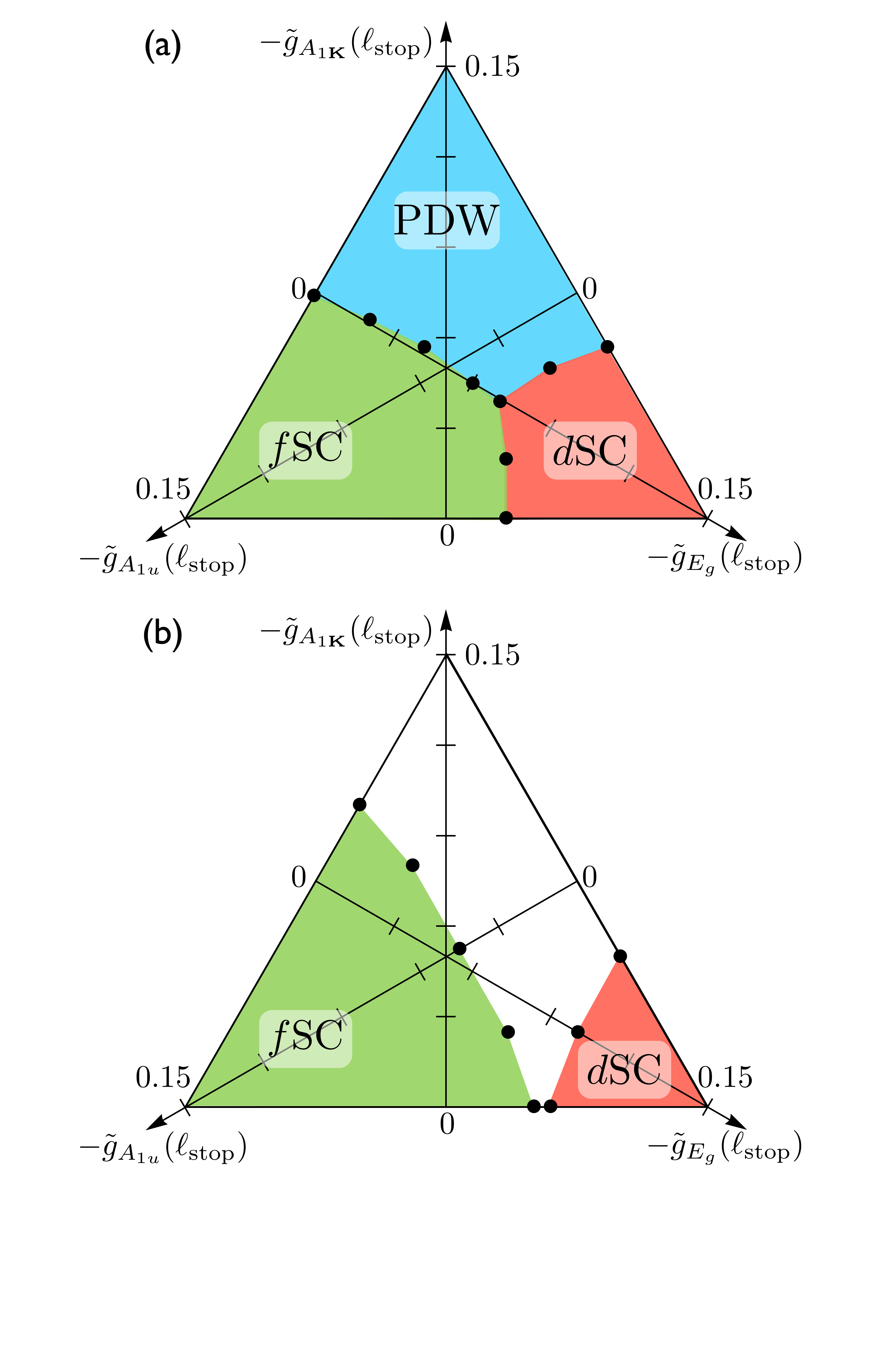}
\caption{Mean field phase diagrams for superconducting phases, in which couplings and other parameters are obtained from running the RG up to $\ell = \ell_\mathrm{stop}$, which is defined such that $\tilde g_{E_g} (\ell_\mathrm{stop}) + \tilde g_{A_{1u}} (\ell_\mathrm{stop}) + \tilde g_{A_{1\mathbf{K}}} (\ell_\mathrm{stop}) = -0.15$. (a) Phase diagram without trigonal warping ($v_3 = 0$), with $\mu(\ell_\mathrm{stop}) = 0.9 \Lambda^2/2m^*$ and $T=0$. (b) Phase diagram with trigonal warping ($v_3 = 2.0 \Lambda/2m^*$), $\mu (\ell_\mathrm{stop}) = 1.1 \Lambda^2/2m^*$, and $T(\ell_\mathrm{stop}) = 10^{-3} \Lambda^2/2m^*$. No ordering occurs in any channel in the white portion of the phase diagram.
\label{fig:mean_field_phases}}
\end{figure}

The results of this analysis are shown in Figure \ref{fig:mean_field_phases}.
The case with $T=v_3=0$ is shown in Figure \ref{fig:mean_field_phases}(a). We see that if one of the couplings $-\tilde g_i(\ell_\mathrm{stop})$ is much larger than the others, then the corresponding superconducting phase is selected from the mean-field analysis. For intermediate values of the couplings, there is a first-order transition between the superconducting phases. In Figure \ref{fig:g_flows0} it was found that $\tilde g_{A_{1\mathbf{K}}}(\ell \lesssim \ell_\mathrm{FS})$ was the most negative coupling for both short- and long-ranged interactions. Therefore, the mean-field calculation indicates that the corresponding superconducting state, which is the PDW, will be realized. This is in agreement with the RG analysis of the susceptibilities. The diagram also indicates that, even if $\tilde g_{E_g}(\ell_\mathrm{stop})$ were slightly greater than the PDW and $f$-wave couplings, as might happen for example once a small $v_3$ is introduced, the system will still prefer to condense into one of the latter two phases. Thus we see that the superconducting phase can not necessarily be determined in all cases simply by taking the largest coupling when the couplings begin to grow large.

Figure  \ref{fig:mean_field_phases}(b) shows the mean-field phase diagram that results from running the RG up to $\ell_\mathrm{stop}$ in the presence of trigonal warping. (In this case, a finite temperature is introduced in order to avoid the singularity associated with integrating through the disconnected portions of the Fermi surface.) In this case, one finds that there is no longer any mean-field solution for the PDW phase (though one might appear for even larger values of $-\tilde g_{A_{1\mathbf{K}}}(\ell_\mathrm{stop})$), due to the fact that---as mentioned previously---there is no Cooper logarithm for this state in the absence of intrapocket $\bk \to -\bk$ symmetry. Furthermore, one finds that the $d$-wave phase can only be realized once $\tilde g_{E_g}(\ell_\mathrm{stop})$ becomes significantly more negative than $\tilde g_{A_{1u}}(\ell_\mathrm{stop})$, and that the $f$-wave phase is preferred when these couplings are comparable, with the order parameters for these phases going continuously to zero as the corresponding couplings decrease in magnitude.

In addition to addressing the competition between various superconducting phases, the mean-field analysis presented here corroborates the results of the two preceding subsections. In particular, the $d$-wave and PDW phases shown in Figure \ref{fig:mean_field_phases} are found to exhibit chiral and non-chiral combinations of the order parameters, respectively, in agreement with the results of the free energy expansion.

\section{Discussion}
\label{sec:discussion}
In this paper we have addressed the consequences of electron-electron interactions as a function of carrier doping using a RG approach that allows for particle-hole and superconducting orders to be treated on equal footing. While this perturbative RG scheme can only be formally justified in the weak-coupling limit, the similarity of the phase diagrams shown in Figure \ref{fig:phases3d} to those of the more strongly correlated materials is suggestive that similar mechanisms may be at play in such systems.\cite{paglione10,armitage10,scalapino12}

As we pointed out in Section \ref{sec:mu} and in our previous work\cite{vafek13}, in the special case $\mu = T = v_3 =0$, the values of the renormalized couplings for any $\ell$ are proportional to the bare coupling magnitude $g$. In particular, the magnitude of an attractive coupling at $\ell = \ell_\mathrm{stop} > \ell_1$ is proportional to the initial {\em repulsive} coupling. This implies that, if superconductivity can be realized, one would expect to have the BCS-type relation $T_c \sim e^{-c_1/|\tilde g_i(\ell_\mathrm{stop})|} \sim e^{-c_2/g}$. This is in contrast to the Kohn-Luttinger result, in which the attraction comes about through second-order perturbation theory, and one always obtains $T_c \sim e^{-c_3/g^2}$, which is parametrically smaller in $g$. (Here $c_i$ are  constants $\sim O(1)$.)

\begin{figure}
\centering
\includegraphics[width=0.4\textwidth]{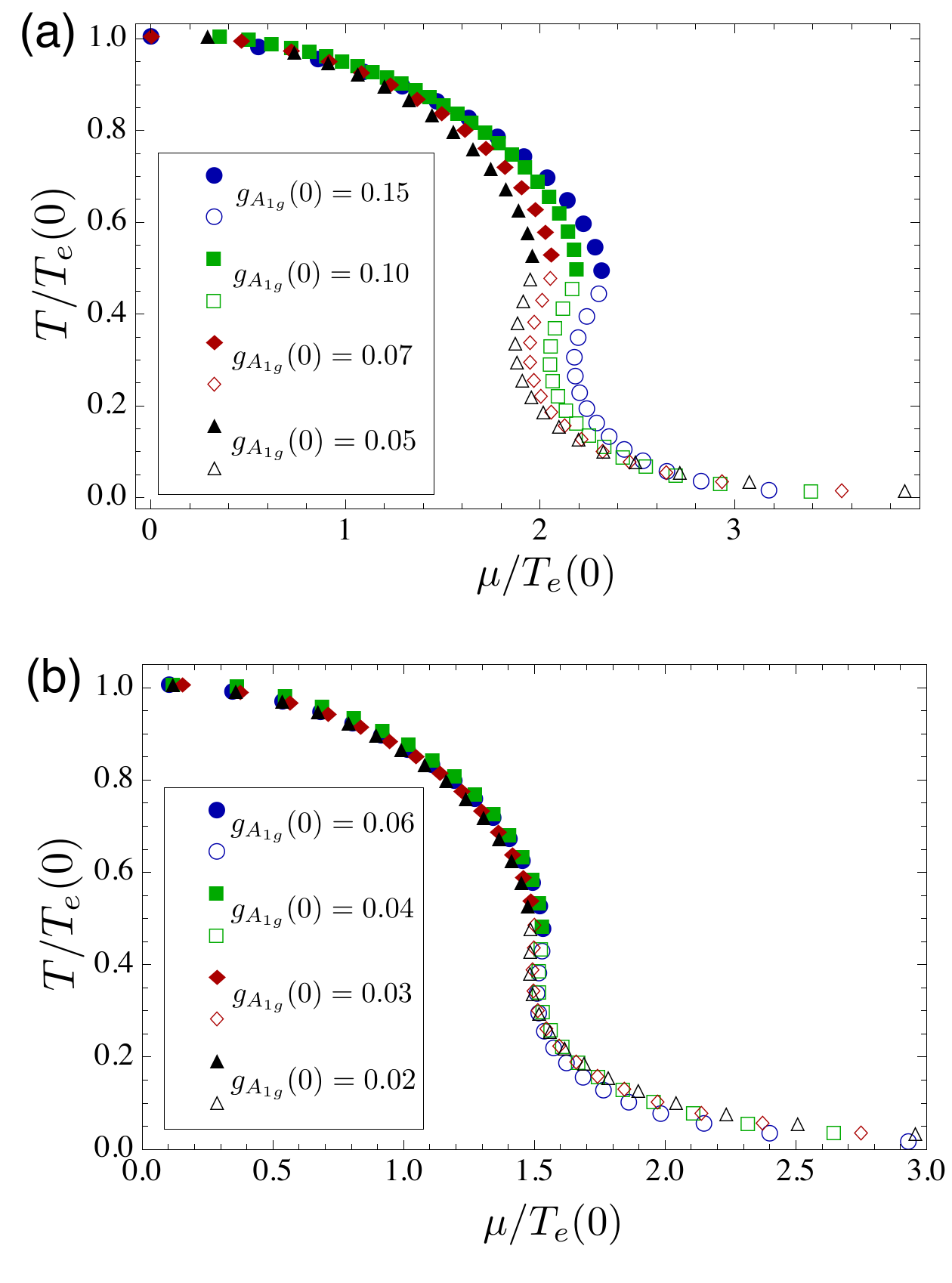}
\caption{Phase instabilities for various values of initial coupling $g_i$, with each axis scaled by the critical temperature for the excitonic phase at $\mu=0$, and $v_3=0$. Closed (open) symbols represent instabilities to excitonic (superconducting) order at temperature $T_e$ ($T_c$). (a) Near-forward scattering interaction, with bare couplings $g_{A_{2u}}(0) = g_{E_\mathbf{K}}(0) = 0.02 g_{A_{1g}}(0)$. (b) Hubbard interaction, with $g_{A_{2u}}(0) = 2g_{E_\mathbf{K}}(0) = g_{A_{1g}}(0)$.
\label{fig:scaling}}
\end{figure}

Of course, the scaling behavior described by \eqref{eq:g_scaling} breaks down once $\mu \neq 0$, which is certainly necessary for obtaining superconductivity. In this case we can show that the above argument remains valid by solving the equations numerically for different values of $g$. Figure \ref{fig:scaling} shows the approximate scaling behavior of the phase instabilities with varying magnitude of the initial coupling.
It is clear from the figure that, regardless of the magnitude of the bare coupling, the maximum critical temperature for superconductivity is in all cases a sizeable fraction of the maximum critical temperature for the excitonic phase, {\em i.e.}\ $T_c^\mathrm{max} \approx 0.4 T_e(\mu=0)$ for any $g$. It is known that $T_e(\mu=0) \sim e^{-c/g}$, where $g$ is the bare coupling strength, and $c \sim O(1)$ is a constant.\cite{cvetkovic12,vafek13} The above observations therefore imply that we also have $T_c^\mathrm{max} \sim e^{-c/g}$.


In Section \ref{sec:temperature} it was shown that, for sufficiently large $\mu / T$, the coupling corresponding to only a single superconducting channel grows to a large value under RG flow, with no competing instabilities in other superconducting or particle-hole channels. This observation lends justification to RG approaches that consider only particle-particle contributions to the flow equations, arguing that---at least for $\mu \gg T$ and away from special fine-tuned nesting conditions---superconductivity is the only generic instability of fermionic systems at finite density.\cite{raghu10,polchinski92,shankar94} However, as shown in Figure \ref{fig:phases3d}, the highest values of superconducting $T_c$ occur near the region of the phase diagram exhibiting a particle-hole instability. Here fluctuations in particle-hole channels clearly play a role in enhancing $T_c$, and the above reasoning, which ignores these fluctuations, breaks down. The enhancement of $T_c$ in this intermediate region relies on the fact that our theory is capturing the crossover between two different dynamical regimes. The first consists of integrating out modes with high energies $E\gg\mu$, in which case the chemical potential plays little role, the scaling behavior is governed by the dynamical critical exponent $z=2$, and fluctuations in both particle-hole and particle-particle channels are comparably important. This regime is where the attractive interaction is generated. In the second regime, the Fermi surface plays a dominant role, constraining the types of scattering processes that are allowed and favoring attractive couplings over repulsive ones at lower energies, thereby leading to superconductivity. The behavior in this regime is similar to that obtained from a $z=1$ theory with linearized fermion dispersion and a cutoff energy near the Fermi surface.\cite{shankar94}
In situations where $\mu$ is comparable to interaction-related energy scales such as $T_e$ and $T_c$, both of the regimes described above are crucial to obtaining the correct physics. In such cases, keeping the Fermi surface as a rigid kinematic constraint may not be the most useful approach, and $\mu$ can instead be treated as a {\em relevant perturbation} away from the charge-neutrality point.

We conclude by discussing the prospects for the experimental observation of the unconventional superconducting phase in doped bilayer graphene. Such observation may be possible, but would be challenging due to the low predicted value of the transition temperature. Obtaining the value of $\mu$ required to induce superconductivity should not present a problem.  As we have shown in Figure \ref{fig:phases3d}, and \ref{fig:scaling}, the critical value $\delta\mu$ required to induce superconductivity is of the same order as the maximum critical temperature for the excitonic phase. Thus we would expect $\delta\mu \sim 1$ meV, or in terms of carrier density away from half filling, $\delta n \approx 10^{10}\sim10^{11}$ cm$^{-2}$, which is well within the resolution of current experiments.\cite{mayorov11} If recent reports \cite{bao12} of a gap at the neutrality point below $T_e \sim 5$ K indeed correspond to an interaction-induced symmetry-breaking phase such as the particle-hole phases described here, then our model would predict that the doped sample should become superconducting below $T_c \sim 1$ K. This value could be further reduced, however, due to the fact that disorder tends to suppress the $T_c$ of unconventional superconducting states such as those considered here. A rough estimate of this suppression may be obtained using the reported \cite{mayorov11} mobilities of \hbox{$\mu \sim 10^6$ cm$^2$ / V s}, which translates to a scattering rate of \hbox{$\tau^{-1} \sim 6\times10^{10}$ s$^{-1}$}, or \hbox{$\hbar \tau^{-1} \sim 0.04$ meV.} Due to the fact that charge carriers can be added electrostatically, similarly high mobilities should be attainable in doped samples. According to the Abrikosov-Gor'kov theory\cite{abrikosov61}, the critical temperature $T_c$ is suppressed to zero when this energy is of the order of the superconducting gap $\Delta_\mathrm{SC}$. If we assume that \hbox{$\Delta_\mathrm{SC} \sim T_c$}, then we have $\Delta_\mathrm{SC} \sim 0.1\ \mathrm{meV} \sim \hbar \tau^{-1}$, so one would expect significant or perhaps complete suppression of $T_c$ in current samples. The above analysis could be further complicated due to fluctuation and finite-size effects, which could lead to further suppression of $T_c$. Thus, while experimental observation of the superconducting state would be a considerable challenge, it may prove to be possible with continued improvement in sample size and quality.

\begin{acknowledgments}
The authors are grateful for helpful discussions with V.\ Cvetkovic and R.\ Throckmorton. 
This work was supported by the NSF CAREER award
under Grant No. DMR-0955561 (OV), NSF Cooperative
Agreement No. DMR-0654118, and the State of Florida
(OV,JM), as well as by ICAM-I2CAM (NSF grant
DMR-0844115), and by DoE, Office of Basic Energy Sciences,
Division of Materials Sciences and Engineering under
Award DE-FG02-08ER46544 (JM).
\end{acknowledgments}
\vspace{1cm}

\appendix
\section{Flow equations for chemical potential and couplings}
\label{sec:flow_coefficients}

In this section we provide the detailed forms of the flow equations for the couplings and chemical potential. We shall make use of the following dimensionless parameters:
\ba
t &= \frac{T}{\Lambda^2 / 2m^*}, \\
\tilde\mu &= \frac{\mu}{\Lambda^2 / 2m^*}, \\
\nu_3 &= \frac{v_3}{\Lambda / 2m^*}.
\ea
In the main text it was noted that the temperature and trigonal warping velocity flow according to $t_\ell = t e^{2\ell}$ and $\nu_{3\ell} = \nu_3 e^\ell$, respectively. The flows for chemical potential and the couplings are more complicated, and we proceed to discuss these below. 

The one-loop flow equation for the total chemical potential, $\mu + \delta\mu = \frac{\Lambda^2}{2m^*}(\tilde\mu + \delta\tilde\mu)$, where $\delta\mu$ is the chemical potential of the half-filled system, comes from evaluating the diagram in Figure \ref{RG}(c) and is given by
\ba
\label{mu'_flow}
\frac{d}{d\ell} (\tilde\mu_\ell + \delta\tilde\mu_\ell)&= 
	2 (\tilde\mu_\ell + \delta\tilde\mu_\ell) \\
	&- 2 [1+\tilde K (\tilde\mu_\ell + \delta\tilde\mu_\ell, t_\ell, \nu_{3\ell})] \sum_i c_i g_i (\ell),
\ea
where
\be
\label{g_sum}
\sum_{i=1}^9 c_i g_i = 8 g_{A_{1g}} - \sum_{j=1}^9 \sum_{m=1}^{m_j} g_j ,
\ee
and
\ba
\label{K_tilde}
\tilde K(\tilde\mu, t, \nu_{3}) = \frac{1}{2\pi} \int_{-1}^1 & \frac{dx}{\sqrt{1-x^2}} \bigg[
 	\tanh \left( \frac{Q_+(\nu_{3}, x) + \tilde\mu}{2 t} \right) \\
	&- \tanh \left( \frac{Q_+(\nu_{3}, x) - \tilde\mu}{2 t} \right) \bigg],
\ea
with
\be
Q_\pm(\nu_3, x) = \sqrt{1 + \nu_3^2 \pm 2 \nu_3 x}. 
\ee
The function appearing in \eqref{K_tilde} is related to $K$ from \eqref{mu_flow} in the main text as $K(\mu,T,v_3) = \frac{\Lambda^2}{2m^*} \tilde K(\tilde\mu,t,\nu_3)$.

Although the chemical potential at half filling, $\delta\mu$, vanishes in the noninteracting system, its value will be shifted in the presence of interactions. It can be computed exactly by carrying out the following particle-hole transformation for the four-component fermionic operators:
\ba
\label{PH}
\psi_{\bk\sigma} &= \tau_1\sigma_3 \chi_{-\bk\sigma}^*, \\
\psi_{\bk\sigma}^* &= \tau_1\sigma_3 \chi_{-\bk\sigma},
\ea 
where, as before, $\tau_i$ and $\sigma_i$ are Pauli matrices operating on valley and layer indices. By rewriting the Hamiltonian given by \eqref{H0} and \eqref{H_int} in terms of these new operators and carefully observing anticommutation relations, one finds that it remains invariant under the transformation \eqref{PH} at $\mu =0$ for 
\be
\label{delta_mu}
\delta\tilde\mu = \sum_{i=1}^9 c_i g_i ,
\ee
where the sum on the right hand side is again given by \eqref{g_sum}. 
The expression \eqref{delta_mu} remains valid when the parameters flow, {\em i.e.}\ for $\delta\mu \to \delta\mu_\ell$ and $g_i \to g_i(\ell)$. With $\delta\mu$ included explicitly in the action \eqref{action}, the half-filled, particle-hole symmetric case is realized at $\mu=0$ for any values of the couplings $g_i$. 

It is useful to rewrite the flow equation \eqref{mu'_flow} in terms of $\tilde\mu$ only, which gives the deviation of the chemical potential away from half filling. First, by taking the derivative of \eqref{delta_mu}, we see that the flow equation for $\delta\tilde\mu_\ell$ can be set to zero at the order to which we are working, due to the fact that $dg_i/d\ell \sim g^2$. Using this fact together with \eqref{delta_mu}, we obtain the flow equation for $\tilde\mu_\ell$:
\be
\label{mu0_flow}
\frac{d \tilde\mu_\ell}{d\ell} = 
	2 \tilde\mu_\ell
	- 2 \tilde K (\tilde\mu_\ell, t_\ell, \nu_{3\ell}) \sum_i c_i g_i (\ell),
\ee
which is valid to leading order in the perturbative expansion. Note that we have not included $\delta\tilde\mu_\ell$ in the first argument of $\tilde K$, due to the fact that $\delta\tilde\mu \sim g_i$, and we can ignore this correction to the flow equation at the order to which we are working. Equation \eqref{mu0_flow} is precisely the flow equation \eqref{mu_flow} from the main text. Note in particular that the second term on the right hand side of \eqref{mu0_flow} vanishes when $T=0$ (so long as the cutoff remains above the Fermi surface), and in this case the chemical potential flows according to its tree level scaling ($\tilde\mu_\ell = \tilde\mu_0 e^{2\ell}$). In addition, we see that $\tilde\mu_\ell$ is not generated at any temperature if it is zero initially, as is indeed required by particle-hole symmetry.

The one-loop flow equations for the couplings are
\begin{widetext}
\be
\label{g_flow0}
\frac{d g_i}{d\ell} = \sum_{j,k=1}^9 \sum_{a=1}^4 g_j g_k 
\left[ \sum_{m=1}^4 A_{ijk}^{(a)} (m) F_{ph}^{(a)} (t_\ell, \tilde \mu_\ell, \nu_{3\ell}) 
	+ A_{ijk}^{(a)}(5) F_{pp}^{(a)} (t_\ell, \tilde \mu_\ell, \nu_{3\ell}) \right].
\ee
Note that the shift $\delta\tilde\mu$ has not been included in the argument of $F_{ph,pp}^{(a)}$ in \eqref{g_flow0}, which is justified at the order to which we are working. The functions $F_{ph,pp}^{(a)}$ come from the loop integrals over fast modes, as shown in Figure \ref{RG}(b):
\ba
\label{gg_tw}
T \sum_n &\int_{\Lambda (1-d\ell)}^{\Lambda} \frac{dk}{2 \pi} k 
	\int_0^{2 \pi} \frac{d \theta_\bk}{2 \pi}
	G_0 (i \omega_n , \bk) \otimes G_0 ( \pm i \omega_n, \pm \bk) \\
	&= \frac{m^*}{8 \pi} d\ell \bigg\{ 
	\mp 1_4 \otimes 1_4 \left[ F_{ph,pp}^{(1)} + F_{ph,pp}^{(2)} \right]
	+\frac{1}{2} (1 \sigma_1  \otimes 1 \sigma_1  
	+ \tau_3 \sigma_2  \otimes \tau_3 \sigma_2 )
	\left[ F_{ph,pp}^{(3)} + F_{ph,pp}^{(4)} \right] \bigg\} \\
	&\quad+ \frac{m^*}{8 \pi} d\ell \bigg\{ 
	-\tau_3 1 \otimes \tau_3 1 
	\left[ F_{ph,pp}^{(1)} - F_{ph,pp}^{(2)} \right]
	\pm \frac{1}{2} 
	(\tau_3 \sigma_1 \otimes \tau_3 \sigma_1 + 1 \sigma_2  \otimes 1 \sigma_2 )
	\left[ F_{ph,pp}^{(4)} - F_{ph,pp}^{(3)} \right] \bigg\},
\ea
where the upper and lower signs correspond to the particle-hole and particle-particle cases, respectively, and we have used the following noninteracting Green function:
\ba
\label{g_tw}
G_0 (i \omega_n, \mathbf{k}) &= \left[ (-i \omega_n - \mu) 1_4 
	+ \frac{1}{2m^*}(k_x^2 - k_y^2) 1 \sigma_1 
	+ \frac{k_x k_y}{m^*} \tau_3 \sigma_2 + v_3 k_x \tau_3 \sigma_1  
	- v_3 k_y 1 \sigma_2  \right] ^{-1} \\ 
&= \frac{1}{2} \sum_{s=\pm} (1 + s \tau_3) 
	\frac{(i \omega_n + \mu) 1 
	+ (\varepsilon_\bk \cos 2 \theta_\bk + s v_3 k \cos \theta_\bk ) \sigma_1
	+ (s \varepsilon_\bk \sin 2 \theta_\bk - v_3 k \sin \theta_\bk ) \sigma_2}
	{-(i \omega_n + \mu)^2 + \varepsilon_\bk^2 + v_3^2 k^2 
	+ 2s \varepsilon_\bk v_3 k \cos 3 \theta_\bk} .
\ea
The $F$ functions in \eqref{gg_tw} are defined as
\ba
\label{bigF}
F_{ph,pp}^{(1)} (t, \tilde \mu, \nu_3) 
	&= \frac{1}{2 \pi t} \int_{-1}^1 \frac{dx}{\sqrt{1-x^2}} 
	\Upsilon_{ph,pp}^{(1)} (t, \tilde\mu, \nu_3, x) \\
F_{ph,pp}^{(2)} (t, \tilde \mu, \nu_3) 
	&= \frac{1}{2 \pi \nu_3} \int_{-1}^1 \frac{dx}{\sqrt{1-x^2}} \frac{1}{x} 
	\Upsilon_{ph,pp}^{(2)} (t, \tilde\mu, \nu_3, x) \\
F_{ph,pp}^{(3)} (t, \tilde \mu, \nu_3) 
	&= \frac{1-\nu_3^2}{2 \pi \nu_3} \int_{-1}^1 \frac{dx}{\sqrt{1-x^2}} \frac{1}{x} 
	\Upsilon_{ph,pp}^{(3)} (t, \tilde\mu, \nu_3, x) \\
F_{ph,pp}^{(4)} (t, \tilde \mu, \nu_3) 
	&= \frac{1}{2 \pi t} \int_{-1}^1 \frac{dx}{\sqrt{1-x^2}} \Upsilon_{ph,pp}^{(4)} (t, \tilde\mu, \nu_3, x),
\ea
where
\ba
\Upsilon_{ph}^{(1)} (t, \tilde\mu, \nu_3, x) &= \frac{1}{2} 
	\left[ \frac{1}{\cosh^2 \left( \frac{Q_+ - \tilde \mu}{2 t} \right)}
	+ \frac{1}{\cosh^2 \left( \frac{Q_+ + \tilde \mu}{2 t} \right)} \right] 
	+\frac{t}{Q_+} \left[ \tanh \left( \frac{Q_+ - \tilde \mu}{2 t} \right)
	+ \tanh \left( \frac{Q_+ + \tilde \mu}{2 t} \right) \right] \\
\Upsilon_{ph}^{(2)} (t, \tilde\mu, \nu_3, x) &=  \frac{1}{2} \sum_{\lambda = \pm} \lambda Q_\lambda
	\left[ \tanh \left( \frac{Q_\lambda - \tilde \mu}{2 t} \right)
	+ \tanh \left( \frac{Q_\lambda + \tilde \mu}{2 t} \right) \right] \\
\Upsilon_{ph}^{(3)} (t, \tilde\mu, \nu_3, x) &= -\frac{1}{2} \sum_{\lambda = \pm} 
	\frac{\lambda}{Q_\lambda} \left[ \tanh \left( \frac{Q_\lambda - \tilde \mu}{2 t} \right)
	+ \tanh \left( \frac{Q_\lambda + \tilde \mu}{2 t} \right) \right] \\
\Upsilon_{ph}^{(4)} (t, \tilde\mu, \nu_3, x) &= - \frac{1}{2} 
	\left[ \frac{1}{\cosh^2 \left( \frac{Q_+ - \tilde \mu}{2 t} \right)}
	+ \frac{1}{\cosh^2 \left( \frac{Q_+ + \tilde \mu}{2 t} \right)} \right]
	+\frac{t}{Q_+} \left[ \tanh \left( \frac{Q_+ - \tilde \mu}{2 t} \right)
	+ \tanh \left( \frac{Q_+ + \tilde \mu}{2 t} \right) \right]
\ea
and
\ba
\Upsilon_{pp}^{(1)} (t, \tilde\mu, \nu_3, x) &= \frac{t}{\tilde \mu} 
	\left[ \frac{Q_+ + 2 \tilde \mu}{Q_+ + \tilde \mu} \tanh \left( \frac{Q_+ + \tilde \mu}{2t} \right)
	- \frac{Q_+ - 2 \tilde \mu}{Q_+ - \tilde \mu} \tanh \left( \frac{Q_+ - \tilde \mu}{2t} \right) \right] \\
\Upsilon_{pp}^{(2)} (t, \tilde\mu, \nu_3, x) &= 2 \nu_3 x \sum_{\lambda=\pm} \bigg[ 
	\frac{Q_\lambda - 2 \tilde \mu}{(Q_\lambda - 2 \tilde \mu)^2 - Q_{-\lambda}^2} 
	\tanh \left( \frac{Q_\lambda - \tilde \mu}{2t} \right) 
	+ \frac{Q_\lambda + 2 \tilde \mu}{(Q_\lambda + 2 \tilde \mu)^2 - Q_{-\lambda}^2} 
	\tanh \left( \frac{Q_\lambda + \tilde \mu}{2t} \right) \bigg] \\
\Upsilon_{pp}^{(3)} (t, \tilde\mu, \nu_3, x) &= -2 \nu_3 x \sum_{\lambda=\pm}\frac{1}{Q_\lambda}
	\bigg[ \frac{1}{(Q_\lambda - 2 \tilde \mu)^2 - Q_{-\lambda}^2} 
	\tanh \left( \frac{Q_\lambda - \tilde \mu}{2t} \right) 
	+ \frac{1}{(Q_\lambda + 2 \tilde \mu)^2 - Q_{-\lambda}^2 } 
	\tanh \left( \frac{Q_\lambda + \tilde \mu}{2t} \right) \bigg] \\
\Upsilon_{pp}^{(4)} (t, \tilde\mu, \nu_3, x) &= \frac{t Q_+}{\tilde \mu} 
	\left[ \frac{1}{Q_+ - \tilde \mu} \tanh \left( \frac{Q_+ - \tilde \mu}{2t} \right)
	- \frac{1}{Q_+ + \tilde \mu} \tanh \left( \frac{Q_+ + \tilde \mu}{2t} \right) \right]. \\
\ea
In the limit of vanishing chemical potential, the functions $F_{ph,pp}^{(i)} (t, \tilde \mu, \nu_3)$ reduce to $\Phi_i (t, \nu_3)$ from Ref.\ \onlinecite{cvetkovic12}.

The explicit expressions for the coefficients $A_{ijk}^{(a)} (m)$ were originally derived in Ref.\ \onlinecite{cvetkovic12} and are provided here for completeness. [Note that there are slight differences between the following expressions and those in Ref. \onlinecite{cvetkovic12} due to the fact that our $\Gamma_i^{(m)}$ are defined as $4\times 4$ (rather than $8\times 8$) matrices.] The coefficients that come from evaluating the first diagram shown in Figure \ref{RG}(b) are
\ba
A_{iii}^{(1/2)}(1)&= -\{4\pm\mbox{Tr}[(\Gamma_i^{(1)}\tau_3 1)^2]\}, \\
A_{iii}^{(3/4)}(1)&= \tfrac{1}{2}\{\mbox{Tr}[(\Gamma_i^{(1)}1\sigma_1)^2]\mp\mbox{Tr}[(\Gamma_i^{(1)}\tau_3\sigma_1)^2] \\
	&\quad\quad\quad\mp\mbox{Tr}[(\Gamma_i^{(1)}1\sigma_2)^2]+\mbox{Tr}[(\Gamma_i^{(1)}\tau_3\sigma_2)^2]\},
\ea
where $\Gamma_i^{(m)}$ are the $4 \times 4$ matrices given in Table \ref{table:matrices}. 
In deriving these coefficients, the completeness relation $\mbox{Tr}(\Gamma_i^{(m)}\Gamma_j^{(n)})=4\delta_{ij}\delta_{mn}$ has been used. The superscripts on the left hand sides of the above equations correspond to the upper and lower signs on the right hand sides. As one would expect for the ``RPA''-type diagram shown in Figure \ref{RG}(b), only terms diagonal in the couplings are nonzero, with $A_{ijk}^{(a)}(1) \sim \delta_{ij}\delta_{jk}$. From the second and third diagrams in Figure \ref{RG}(b), the nonzero contributions to the $A_{ijk}^{(a)}$ coefficients are
\ba
A_{iij}^{(1/2)}(2+3)&=\tfrac{1}{4}\sum_{m=1}^{m_j}\{\mbox{Tr}[(\Gamma_i^{(1)}\Gamma_j^{(m)})^2]\pm\mbox{Tr}(\Gamma_i^{(1)}\Gamma_j^{(m)}\tau_3 1\Gamma_i^{(1)}\tau_3 1\Gamma_j^{(m)})\}, \\
A_{iij}^{(3/4)}(2+3)&=\tfrac{-1}{8}\sum_{m=1}^{m_j}[\mbox{Tr}(\Gamma_i^{(1)}\Gamma_j^{(m)}1\sigma_1\Gamma_i^{(1)}1\sigma_1\Gamma_j^{(m)})\mp\mbox{Tr}(\Gamma_i^{(1)}\Gamma_j^{(m)}\tau_3\sigma_1\Gamma_i^{(1)}\tau_3\sigma_1\Gamma_j^{(m)}) \cr
&\quad\quad\quad\mp\mbox{Tr}(\Gamma_i^{(1)}\Gamma_j^{(m)}1\sigma_2\Gamma_i^{(1)}1\sigma_2\Gamma_j^{(m)})+\mbox{Tr}(\Gamma_i^{(1)}\Gamma_j^{(m)}\tau_3\sigma_2\Gamma_i^{(1)}\tau_3\sigma_2\Gamma_j^{(m)})].
\ea
From the fourth diagram in Figure \ref{RG}(b),
\ba
A_{kij}^{(1/2)}(4)&=\tfrac{1}{32}\sum_{m=1}^{m_i}\sum_{n=1}^{m_j}[\mbox{Tr}(\Gamma_k^{(1)}\Gamma_i^{(m)}\Gamma_j^{(n)})\mbox{Tr}(\Gamma_k^{(1)}\Gamma_j^{(n)}\Gamma_i^{(m)})\\
&\quad\quad\quad\quad\pm\mbox{Tr}(\Gamma_k^{(1)}\Gamma_i^{(m)}\tau_3 1\Gamma_j^{(n)})\mbox{Tr}(\Gamma_k^{(1)}\Gamma_j^{(n)}\tau_3 1\Gamma_i^{(m)})], \\
A_{kij}^{(3/4)}(4)&=\tfrac{-1}{64}\sum_{m=1}^{m_i}\sum_{n=1}^{m_j}[\mbox{Tr}(\Gamma_k^{(1)}\Gamma_i^{(m)}1\sigma_1\Gamma_j^{(n)})\mbox{Tr}(\Gamma_k^{(1)}\Gamma_j^{(n)}1\sigma_1\Gamma_i^{(m)})\\
&\quad\quad\quad\quad\mp\mbox{Tr}(\Gamma_k^{(1)}\Gamma_i^{(m)}\tau_3\sigma_1\Gamma_j^{(n)})\mbox{Tr}(\Gamma_k^{(1)}\Gamma_j^{(n)}\tau_3\sigma_1\Gamma_i^{(m)}) \cr
&\quad\quad\quad\quad\mp\mbox{Tr}(\Gamma_k^{(1)}\Gamma_i^{(m)}1\sigma_2\Gamma_j^{(n)})\mbox{Tr}(\Gamma_k^{(1)}\Gamma_j^{(n)}1\sigma_2\Gamma_i^{(m)})\\
&\quad\quad\quad\quad+\mbox{Tr}(\Gamma_k^{(1)}\Gamma_i^{(m)}\tau_3\sigma_2\Gamma_j^{(n)})\mbox{Tr}(\Gamma_k^{(1)}\Gamma_j^{(n)}\tau_3\sigma_2\Gamma_i^{(m)})]. \\
\ea
Finally, from the fifth (particle-particle) diagram,
\ba
A_{kij}^{(1/2)}(5)&=-\tfrac{1}{32}\sum_{m=1}^{m_i}\sum_{n=1}^{m_j}\{[\mbox{Tr}(\Gamma_k^{(1)}\Gamma_i^{(m)}\Gamma_j^{(n)})]^2\mp[\mbox{Tr}(\Gamma_k^{(1)}\Gamma_i^{(m)}\tau_3 1\Gamma_j^{(n)})]^2\}, \\
A_{kij}^{(3/4)}(5)&=-\tfrac{1}{64}\sum_{m=1}^{m_i}\sum_{n=1}^{m_j}\{[\mbox{Tr}(\Gamma_k^{(1)}\Gamma_i^{(m)}1\sigma_1\Gamma_j^{(n)})]^2\pm[\mbox{Tr}(\Gamma_k^{(1)}\Gamma_i^{(m)}\tau_3\sigma_1\Gamma_j^{(n)})]^2 \cr
&\quad\quad\quad\quad\quad\quad\pm[\mbox{Tr}(\Gamma_k^{(1)}\Gamma_i^{(m)}1\sigma_2\Gamma_j^{(n)})]^2+[\mbox{Tr}(\Gamma_k^{(1)}\Gamma_i^{(m)}\tau_3\sigma_2\Gamma_j^{(n)})]^2\}. \\
\ea
\end{widetext}
Together, \eqref{mu0_flow} and \eqref{g_flow0} form a set of 10 coupled, first-order differential equations, which can be solved numerically for the $\ell$-dependent chemical potential and couplings.

\section{Asymptotic behavior of couplings and chemical potential}
\label{sec:bilayer_g_mu_flows}
We proceed to analyze the $\ell \to \infty$ behavior of the flow equations for the chemical potential and couplings at the critical temperature. The critical behavior is determined by the asymptotic limit of the flow equations as $\ell \to \infty$. From \eqref{mu'_flow}, the flow equation for the chemical potential becomes
\be
\label{mu'_flow_infty}
\frac{d \tilde\mu_\ell}{d\ell} \stackrel{(\ell \to \infty)}{=} 2 \tilde\mu_\ell
	- 2 \tanh \left( \frac{e^{-2\ell} \tilde\mu_\ell}{2 t_0} \right) \sum_i c_i g_i (\ell).
\ee
As $\ell \to \infty$, the flow equations \eqref{g_flow0} and \eqref{mu'_flow_infty} admit solutions of the form
\be
\label{mu_cases}
\tilde\mu_\ell \sim e^{\alpha \ell}, \quad 
\begin{cases}
\alpha < 2, \quad \sum_i c_i g_i (\ell \to \infty) > 0, \\
\alpha > 2, \quad \sum_i c_i g_i (\ell \to \infty) < 0,
\end{cases}
\ee
with the diverging couplings behaving as $g_i (\ell) \sim e^{2\ell}$ and $g_i(\ell) \sim e^{\alpha \ell}$ for $\alpha < 2$ and $\alpha > 2$, respectively. Below we consider the cases $\alpha < 2$ and $\alpha > 2$ separately.

For $\alpha < 2$, assuming (to be verified self-consistently below) that $\tilde\mu_\ell \sim e^{\alpha \ell}$, the flow equation coefficients in \eqref{bigF} reduce to
\ba
\label{f_infty1}
&F_{ph}^{(1,2)}(\ell \to \infty) = F_{pp}^{(1,2)}(\ell \to \infty) = \frac{1}{t_\ell} \sim e^{-2\ell}, \\
&F_{ph}^{(3,4)}(\ell \to \infty) \approx 0 \approx F_{pp}^{(3,4)}(\ell \to \infty),
\ea
where the functions in the second line vanish faster than $e^{-2\ell}$ and so can be neglected in the limit $\ell \to \infty$. The flow equation for the coupling constants \eqref{g_flow2} then becomes
\ba
\label{g_flow1}
\frac{dg_i}{d\ell} &\stackrel{(\ell \to \infty)}{=} \frac{e^{-2\ell}}{t_0} 
	\sum_{j,k=1}^9 \sum_{m=1}^5 
	\left[ A_{ijk}^{(1)}(m) + A_{ijk}^{(2)}(m) \right] g_j g_k \\
&\equiv e^{-2\ell} \sum_{j,k = 1}^9 \tilde A^{(1)}_{ijk} g_j g_k
\ea
From this we see that the asymptotic behavior of the runaway couplings is $g_i (\ell\to\infty) \sim e^{2\ell}.$ In order to be more concrete, we define the ``coupling magnitude'' as
\be
\label{coupling_mag}
G (\ell) = \sqrt{\sum_{i=1}^9 g_i^2 (\ell)}.
\ee
The flow equation for this quantity is then 
\be
\frac{dG}{d\ell} \stackrel{(\ell \to \infty)}{=} e^{-2\ell} G^2 \sum_{ijk} \tilde A^{(1)}_{ijk} \rho_i \rho_j \rho_k,
\ee
where 
\be
\rho_i \equiv \lim_{\ell \to \infty} \frac{g_i(\ell)}{G(\ell)} = const. 
\ee
We refer to $\rho_i$ as a ``fixed ratio,'' and, as we shall see below when calculating susceptibilities, the relative values of these 9 quantities ultimately determine the nature of the phase instability. The solution to the flow equation for $G(\ell)$ is
\be
\label{big_g}
G (\ell) = \frac{2 e^{2\ell}}{\sum_{ijk} \tilde A^{(1)}_{ijk} \rho_i \rho_j \rho_k}
\ee
The flow equation \eqref{mu'_flow_infty} for $\tilde\mu_\ell$ meanwhile becomes
\ba
\label{mu'_flow1}
\frac{d \tilde\mu_\ell}{d\ell} \stackrel{(\ell \to \infty)}{=} 
	&\left[ 2 - \frac{1}{t_0} \sum_i c_i g_i (\ell) e^{-2\ell} \right] \tilde\mu_\ell  \\
= 2 &\left[ 1 - \frac{\sum_i c_i \rho_i}{t_0 \sum_{ijk} \tilde A_{ijk}^{(1)} \rho_i \rho_j \rho_k}  \right] \tilde\mu_\ell , \quad (\alpha < 2),
\ea
where we have utilized \eqref{big_g} in the second line. Since the quantity in brackets approaches a constant as $\ell \to \infty$, this equation indeed has the solution $\tilde\mu_\ell \sim e^{\alpha \ell}$. 

For $\alpha > 2$, assuming from \eqref{mu_cases} that $\tilde\mu_\ell \sim e^{\alpha \ell}$, the limiting behavior of the functions in \eqref{bigF} is 
\ba
\label{f_infty2}
&F_{pp}^{(1,2)}(\ell \to \infty) = \frac{2}{\tilde\mu_\ell} \sim e^{-\alpha \ell}, \\
&F_{ph}^{(1,2,3,4)}(\ell \to \infty) \approx 0 \approx F_{pp}^{(3,4)}(\ell \to \infty).
\ea
The functions in the second line of \eqref{f_infty2} vanish exponentially faster than $\sim e^{-\alpha \ell}$ as $\ell\to\infty$. In this case the asymptotic behavior of the flow equation for the coupling constants is
\ba
\label{g_flow2}
\frac{dg_i}{d\ell} &\stackrel{(\ell \to \infty)}{=} \frac{2}{\tilde\mu_\ell} \sum_{j,k=1}^9 
	\left[ A_{ijk}^{(1)}(5) + A_{ijk}^{(2)}(5) \right] g_j g_k \\
&\equiv e^{-\alpha \ell} \sum_{j,k = 1}^9 \tilde A^{(2 )}_{ijk} g_j g_k.
\ea
The flow equation for $G(\ell)$ then becomes
\be
\label{G_flow_alpha}
\frac{dG}{d\ell} = e^{-\alpha \ell} G^2 \sum_{ijk} \tilde A^{(2)}_{ijk} \rho_i \rho_j \rho_k,
\ee
which has the solution
\be
\label{big_g_alpha}
G (\ell) = \frac{\alpha e^{\alpha \ell}}{\sum_{ijk} \tilde A^{(2)}_{ijk} \rho_i \rho_j \rho_k}.
\ee
The flow equation \eqref{mu'_flow_infty} in this case reduces to
\ba
\label{mu'_flow2}
\frac{d \tilde\mu_\ell}{d\ell} \stackrel{(\ell \to \infty)}{=} 
	& 2 \tilde\mu_\ell - 2 \sum_i c_i g_i (\ell) \\
= & 2 \tilde\mu_\ell - \frac{2 \alpha e^{\alpha \ell} \sum_i c_i \rho_i}{\sum_{ijk} \tilde A_{ijk}^{(2)} \rho_i \rho_j \rho_k}, \quad (\alpha > 2).
\ea

Putting together \eqref{mu'_flow1} and \eqref{mu'_flow2}, we have in both cases $\tilde\mu_\ell \sim e^{\alpha \ell}$, with
\ba
\label{alpha}
\alpha = \begin{cases}
2 - \frac{2 \sum c_i \rho_i}{t_0 \sum_{ijk} \tilde A_{ijk}^{(1)} \rho_i \rho_j \rho_k} &, 
	\sum_i c_i \rho_i > 0\\
\left[ \frac{1}{2} + \frac{\sum_i c_i \rho_i}{\sum_{ijk} \tilde A_{ijk}^{(2)} \rho_i \rho_j \rho_k} \right]^{-1} &,
	\sum_i c_i \rho_i < 0.
\end{cases}
\ea
We see from \eqref{alpha} that, indeed, $\alpha < 2$ ($\alpha > 2$) when flowing toward a stable ray that satisfies $\sum_i c_i \rho_i > 0$ ($\sum_i c_i \rho_i > 0$), consistent with our initial assumption. Note that, for $\alpha < 2$, the chemical potential does not enter into the asymptotic flow equation for the coupling constants \eqref{g_flow1}, so that the asymptotic analysis presented here matches exactly that from Ref.\ \onlinecite{cvetkovic12}. This means that, although the flow behavior at small $\ell$ may determine which stable ray is approached, the particular ratios that define that ray, as well as the universal properties such as critical exponents associated with it, are independent of the chemical potential. On the other hand, for $\alpha > 2$, the temperature doesn't appear at all in the asymptotic analysis, and the $\ell \to \infty$ behavior depends entirely on the chemical potential. 

\section{Susceptibilities and symmetry breaking}
\label{sec:bilayer_susceptibilities}
The coefficients in the vertex flow equations  \eqref{vertex_flow} come from evaluating the diagrams in Figure \ref{RG}(d). The particle-hole coefficients are given by 
\be
B_{ij}^{ph}(t,\tilde\mu,\nu_3) = \sum_{a=1}^{4} \sum_{m=1}^2 
	B_{ij}^{(a)}(m) F_{ph}^{(a)}(t,\tilde\mu,\nu_3),
\ee
where, from the first diagram,
\begin{widetext}
\ba
\label{b_traces}
B^{(1/2)}_{ij}(1)&=-\tfrac{1}{2}\sum_{n=1}^{m_j}[\mbox{Tr}(O_{ph}^{(i)}(\Gamma_j^{(n)}1))\pm\mbox{Tr}(\tau_3 1_4O_{ph}^{(i)}\tau_3 1_4(\Gamma_j^{(n)}1))], \\
B^{(3/4)}_{ij}(1)&=\tfrac{1}{4}\sum_{n=1}^{m_j}[\mbox{Tr}(1\sigma_11O_{ph}^{(i)}1\sigma_11(\Gamma_j^{(n)}1))\mp\mbox{Tr}(\tau_3\sigma_11O_{ph}^{(i)}\tau_3\sigma_11(\Gamma_j^{(n)}1)) \cr
&\mp\mbox{Tr}(1\sigma_21O_{ph}^{(i)}1\sigma_21(\Gamma_j^{(n)}1))+\mbox{Tr}(\tau_3\sigma_21O_{ph}^{(i)}\tau_3\sigma_21(\Gamma_j^{(n)}1))],
\ea
and from the second diagram,
\ba
B^{(1/2)}_{ij}(2)&=\tfrac{1}{16}\sum_{n=1}^{m_j}\{\mbox{Tr}[(O_{ph}^{(i)}(\Gamma_j^{(n)}1))^2]\pm\mbox{Tr}(O_{ph}^{(i)}(\Gamma_j^{(n)}1)\tau_3 1_4O_{ph}^{(i)}\tau_3 1_4(\Gamma_j^{(n)}1))\}, \\
B^{(3/4)}_{ij}(2)&=-\tfrac{1}{32}\sum_{n=1}^{m_j}[\mbox{Tr}(O_{ph}^{(i)}(\Gamma_j^{(n)}1)1\sigma_11O_{ph}^{(i)}1\sigma_11(\Gamma_j^{(n)}1))\mp\mbox{Tr}(O_{ph}^{(i)}(\Gamma_j^{(n)}1)\tau_3\sigma_11O_{ph}^{(i)}\tau_3\sigma_11(\Gamma_j^{(n)}1)) \cr
&\mp\mbox{Tr}(O_{ph}^{(i)}(\Gamma_j^{(n)}1)1\sigma_21O_{ph}^{(i)}1\sigma_21(\Gamma_j^{(n)}1))+\mbox{Tr}(O_{ph}^{(i)}(\Gamma_j^{(n)}1)\tau_3\sigma_21O_{ph}^{(i)}\tau_3\sigma_21(\Gamma_j^{(n)}1))].
\ea 
The coefficients for the particle-particle vertex flow equations come from evaluating the third diagram in Figure \ref{RG}(d):
\be
B_{ij}^{pp}(t,\tilde\mu,\nu_3) = \sum_{a=1}^{4} C_{ij}^{(a)} F_{pp}^{(a)}(t,\tilde\mu,\nu_3),
\ee
where
\ba
\label{c_traces}
C^{(1/2)}_{ij}&=-\tfrac{1}{16}\sum_{n=1}^{m_j}\{\mbox{Tr}[O_{pp}^{(i)}(\Gamma_j^{(n)}1)O_{pp}^{(i)}(\Gamma_j^{(n)}1)^T]\mp\mbox{Tr}[O_{pp}^{(i)}(\Gamma_j^{(n)}1)\tau_3 1_4 O_{pp}^{(i)}\tau_3 1_4(\Gamma_j^{(n)}1)^T]\},\\
C^{(3/4)}_{ij}&=-\tfrac{1}{32}\sum_{n=1}^{m_j}\{\mbox{Tr}[O_{pp}^{(i)}(\Gamma_j^{(n)}1)1\sigma_11O_{pp}^{(i)}1\sigma_11(\Gamma_j^{(n)}1)^T]\pm\mbox{Tr}[O_{pp}^{(i)}(\Gamma_j^{(n)}1)\tau_3\sigma_11O_{pp}^{(i)}\tau_3\sigma_11(\Gamma_j^{(n)}1)^T] \cr
&\mp\mbox{Tr}[O_{pp}^{(i)}(\Gamma_j^{(n)}1)1\sigma_21O_{pp}^{(i)}1\sigma_21(\Gamma_j^{(n)}1)^T]-\mbox{Tr}[O_{pp}^{(i)}(\Gamma_j^{(n)}1)\tau_3\sigma_21O_{pp}^{(i)}\tau_3\sigma_21(\Gamma_j^{(n)}1)^T]\}.
\ea
\end{widetext}
The $8\times 8$ matrices $O_{ph,pp}^{(i)}$ appearing in \eqref{b_traces} and \eqref{c_traces} are given by
\ba
O_{ph}^{(i)} &= \begin{cases}
	\Gamma_i^{(1)} 1,\quad\quad & 1\leq i \leq 9, \\
	\Gamma_i^{(1)} \sigma_3, & 10 \leq i \leq 18.
	\end{cases} \\
O_{pp}^{(i)} &= \begin{cases}
	\tilde\Gamma_i^{(1)} \sigma_2,\quad\quad & i=1,3,5,7,9, \\
	\tilde\Gamma_i^{(1)} 1, & i=2,4,6,8.
	\end{cases} \\
\ea
As with the other flow equations, the asymptotic behavior of \eqref{vertex_flow} must be analyzed separately for the cases $\alpha < 2$ and $\alpha > 2$.

For $\alpha < 2$, using \eqref{f_infty1} the vertex flow equations become
\ba
\label{vertex_flow_infty1}
\frac{d \ln \Delta_i^{ph}}{d\ell} & \stackrel{(\ell \to \infty)}{=} 2 + 
	\frac{1}{t_\ell} \sum_{j=1}^9 \tilde B_{ij} g_j (\ell), \\
\frac{d \ln \Delta_i^{pp}}{d\ell} & \stackrel{(\ell \to \infty)}{=} 2 +
	\frac{1}{t_\ell} \sum_{j=1}^9 \tilde C_{ij} g_j (\ell).
\ea
where
\ba
\tilde B_{ij} &= \sum_{m=1}^2 [B^{(1)}_{ij}(m) + B^{(2)}_{ij}(m)] \\
\tilde C_{ij} &= [C^{(1)}_{ij} + C^{(2)}_{ij}].
\ea
Note that these results are again independent of the chemical potential, so that the universal features of the critical behavior are identical to those established by Ref.\ \onlinecite{cvetkovic12} when $\alpha < 2$. For $\alpha > 2$, on the other hand, using \eqref{f_infty2} the vertex flow equations become
\ba
\label{vertex_flow_infty2}
\frac{d \ln \Delta_i^{ph}}{d\ell} & \stackrel{(\ell \to \infty)}{=} 2 , \\
\frac{d \ln \Delta_i^{pp}}{d\ell} & \stackrel{(\ell \to \infty)}{=} 
	2 + \frac{2}{\tilde\mu_\ell} \sum_{j=1}^9 \tilde C_{ij} g_j (\ell).
\ea

Putting these results together, and using \eqref{big_g} and \eqref{big_g_alpha}, the vertex flow equations can be expressed as
\be
\frac{d \ln \Delta_i^{ph,pp}}{d\ell} & \stackrel{(\ell \to \infty)}{=} 
\begin{cases} 
2 + \eta_i^{ph,pp} ,  & \alpha < 2, \\
2 + \alpha \eta_i^{ph,pp}/2 ,  & \alpha > 2,
\end{cases}
\ee
with the anomalous critical exponents given by
\be
\label{eta_ph}
\eta_i^{ph} = 
\begin{cases}
\frac{2 \sum_j \tilde B_{ij} \rho_j}{\sum_{ijk} \tilde A^{(1)}_{ijk} \rho_i \rho_j \rho_k}, & \alpha < 2, \\
0, & \alpha > 2,
\end{cases}
\ee
\be
\label{eta_pp}
\eta_i^{pp} = \begin{cases}
\frac{2 \sum_j \tilde C_{ij} \rho_j}{\sum_{ijk} \tilde A^{(1)}_{ijk} \rho_i \rho_j \rho_k}, & \alpha < 2 \\
\frac{2 \sum_j \tilde C_{ij} \rho_j}{\sum_{ijk} \tilde A^{(2)}_{ijk} \rho_i \rho_j \rho_k}, & \alpha > 2.
\end{cases}
\ee
It can be shown following the method of Ref.\ \onlinecite{cvetkovic12} that, as the critical temperature is approached, the corresponding susceptibilities behave as
\be
\chi_i^{ph,pp} \sim (t_0 - t_c)^{-\gamma_i^{ph,pp}},
\ee
with
\be
\gamma_i^{ph,pp} = \eta_i^{ph,pp} - 1.
\ee
Thus the condition for a diverging susceptibility is that $\eta_i^{ph,pp} > 1$, with mean-field behavior realized for \hbox{$\eta_i^{ph,pp} = 2$.}

Due to the fact that $\eta_i^{ph} = 0$ when $\alpha > 2$,  there can be no instability in any particle-hole channel in this case. Thus {\em only} particle-particle instabilities may occur when $\alpha > 2$, {\em i.e.}\ when the flow of the chemical potential under RG is more relevant than the flow of temperature. This could in fact be inferred already from \eqref{g_flow2}, which shows that only the particle-particle ladder diagrams contribute to the asymptotic flows for $\alpha > 2$.

\bibliographystyle{apsrev}

\begin{thebibliography}{10}
\bibitem{SCfootnote} We note that the symmetry-allowed term $\sim \Delta_s^* 
	(|\Delta_{d1}|^2 \Delta_{d1} - 2|\Delta_{d2}|^2 \Delta_{d1} - \Delta_{d1}^* \Delta_{d2}^2) + c.c.$ is also present in general, where $\Delta_s$ is the $A_{1g}$ order parameter. Naively, the presence of such a term in the free energy would imply that $s$-wave superconductivity is immediately induced once one of the $d$-wave components becomes nonzero. However, closer inspection reveals that this term vanishes in the case of chiral $d$-wave superconductivity, where $\Delta_{d2} = e^{\pm i \pi /2}\Delta_{d1}$, so that $s$-wave superconductivity is not induced at $T=T_c$.
\end{thebibliography}

\end {document}